\documentclass[
	twocolumn,
	amsmath,
	amssymb,
	pra,
	superscriptaddress]
{revtex4}
\usepackage{graphicx}
\usepackage{amsmath,amsfonts,amssymb,amsthm}
\usepackage{fancyhdr}
\usepackage{mathrsfs}
\usepackage{epstopdf}
\usepackage{setspace}
\usepackage{bm}
\usepackage[colorlinks]{hyperref}
\usepackage{siunitx}

\begin{document}

\title{Quantum simulation of the spin-boson model with a microwave circuit}

\author{Juha Lepp\"akangas}
\affiliation{Institut f\"ur Theoretische Festk\"orperphysik, Karlsruhe Institute of Technology, 76131  Karlsruhe, Germany}
\affiliation{Physikalisches Institut, Karlsruhe Institute of Technology, 76131 Karlsruhe, Germany}

\author{Jochen Braum\"uller}
\affiliation{Physikalisches Institut, Karlsruhe Institute of Technology, 76131 Karlsruhe, Germany}

\author{Melanie Hauck}
\affiliation{Institut f\"ur Theoretische Festk\"orperphysik, Karlsruhe Institute of Technology, 76131 Karlsruhe, Germany}

\author{Jan-Michael Reiner}
\affiliation{Institut f\"ur Theoretische Festk\"orperphysik, Karlsruhe Institute of Technology, 76131 Karlsruhe, Germany}

\author{Iris Schwenk}
\affiliation{Institut f\"ur Theoretische Festk\"orperphysik, Karlsruhe Institute of Technology, 76131 Karlsruhe, Germany}

\author{Sebastian Zanker}
\affiliation{Institut f\"ur Theoretische Festk\"orperphysik, Karlsruhe Institute of Technology, 76131  Karlsruhe, Germany}

\author{Lukas Fritz}
\affiliation{Institut f\"ur Theoretische Festk\"orperphysik, Karlsruhe Institute of Technology, 76131 Karlsruhe, Germany}

\author{Alexey V. Ustinov}
\affiliation{Physikalisches Institut, Karlsruhe Institute of Technology, 76131 Karlsruhe, Germany}
\affiliation{Russian Quantum Center, National University of Science and Technology MISIS, 119049 Moscow, Russia}

\author{Martin Weides}
\affiliation{Physikalisches Institut, Karlsruhe Institute of Technology, 76131 Karlsruhe, Germany}
\affiliation{Physikalisches Institut, Johannes Gutenberg University Mainz, 55128 Mainz, Germany}

\author{Michael Marthaler}
\affiliation{Institut f\"ur Theoretische Festk\"orperphysik, Karlsruhe Institute of Technology, 76131 Karlsruhe, Germany}
\affiliation{Institut f\"ur Theorie der Kondensierten Materie, Karlsruhe Institute of Technology, 76131 Karlsruhe, Germany}
\affiliation{ Theoretische Physik, Universit\"at des Saarlandes, 66123 Saarbr\"ucken, Germany}



\begin{abstract}

We consider superconducting circuits for the purpose of simulating the spin-boson model.
The spin-boson model consists of a single two-level system coupled to bosonic modes.
In most cases, the model is considered in a limit where the bosonic modes are sufficiently dense to form a continuous spectral bath.
A very well known case is the ohmic bath, where the density of states grows linearly with the frequency.
In the limit of weak coupling or large temperature, this problem can be solved numerically.
If the coupling is strong, the bosonic modes can become sufficiently excited to make a classical simulation impossible.
Here, we discuss how a quantum simulation of this problem can be performed by coupling a superconducting qubit to a set of microwave resonators. 
We demonstrate a possible implementation of a continuous spectral bath with individual bath resonators coupling strongly to the qubit.
Applying a microwave drive scheme potentially allows us to access the strong-coupling regime of the spin-boson model. 
We discuss how the resulting spin relaxation dynamics with different initialization conditions can be probed by
standard qubit-readout techniques from circuit quantum electrodynamics.

\end{abstract}

\maketitle


\section{Introduction}

The spin-boson model studies dynamics of a two-level system interacting with a bosonic environment~\cite{Spin_Boson_Rev,Book_Weiss}.
It is a generic model of quantum decoherence of two-level systems~\cite{Shnirman2002}
and is of particular interest for the studies of quantum phase transitions~\cite{Orth}.
It assumes a linear coupling between a two-level system (spin operator) and a collective coordinate of the bosonic bath.
Despite of its very simple form, the spin-boson model is not exactly solvable by any known theoretical method~\cite{Book_Weiss}.

Certain limits of the spin-boson problem are however well understood. In the limit of weak system-bath coupling,
perturbative methods such as the Born-Markov master equation~\cite{Spin_Boson_Rev,Book_Weiss} can be applied, describing weakly damped coherent oscillations.
In the limit of high temperature,
adequate perturbation theory may be possible in the polaron basis~\cite{Spin_Boson_Rev,Book_Weiss,Marthaler2016}, describing incoherent hopping of dressed states.
In such situations, the corresponding spin-boson model can be solved in a good approximation, analytically or numerically.
On the other hand, when interaction strengths are of the order of involved frequencies,
the problem becomes increasingly difficult, or even impossible, to solve in a desired accuracy.
This regime covers many interesting problems of many-body physics, such as the Kondo effect~\cite{Spin_Boson_Rev,KondoMicrowaves1,KondoMicrowaves2} and localization-delocalization transitions of spin dynamics in different environments~\cite{Spin_Boson_Rev,Bulla2007,Orth}.

A commonly used strategy of obtaining new insight into many quantum models, 
or to test previous theoretical predictions, is the approach of quantum simulation~\cite{Quantum_Simulation_Rev,QuantumSimulation,Porras2007,Schneider2012,Lemmer2017}. The Hamiltonian of the
problem is mapped to a well-controlled artificial quantum system and its dynamics is probed  experimentally.
Superconducting microwave circuits have proven to be a particularly attractive experimental platform for engineering various interesting Hamiltonians~\cite{PhotosynthesisSimulation,PolaronSimulation,FermiHubbardSimulation,FermionFermionScattering,Ballester2012,Silveri2013,Jochen2017,Gross2015} due to its good controllability and feasibility of realizing exotic parameter regimes~\cite{cQED1,cQED2,cQED3,cQED4,Gu2017}.

The computational complexity of model Hamiltonians is connected to the mutual coupling strengths of the individual elements relative to the subsystem energies.
Reaching the strong-coupling regime between a qubit and a resonator in a superconducting microwave circuit, described by the Jaynes-Cummings model, has enabled the reproduction of many fundamental phenomena from cavity quantum electrodynamics (QED) and has led to the development of novel quantum systems and applications~\cite{Blais2004,Wallraff2008,Hofheinz2009,cQED1,cQED2,cQED3,cQED4}. Here, the coupling strength between the qubit and the bosonic mode is larger than the decay rates of the two coupled systems.
If the coupling strength becomes comparable to the sub-system energies, the counter-rotating terms of the general quantum Rabi model cannot be neglected. This ultra-strong coupling regime~\cite{Gunther2009,Anappera2009,Casanova2010}
has been experimentally demonstrated with superconducting circuits~\cite{Forn-Diaz2010,Niemczyk2010,Baust2016,Yoshihara2016,Chen2017,Bosman2017,Roch2018} and in various other platforms \cite{Gerritsma2010,Geiser2012,Maissen2014}.
Interesting phenomena that emerge include
ground-state squeezing~\cite{Ciuti2005}, single-mode phase transitions~\cite{Hwang2015},
and non-classical state generation~\cite{Ashhab2010,Leppakangas2018,Kockum2017}.

The spin-boson model is a generalization of the single-mode quantum Rabi model to a continuous-mode environment.
Near the coupling regime that exhibits Kondo physics and localization-delocalization transitions~\cite{Spin_Boson_Rev,Bulla2007,Orth},
the energy decay rate $\Gamma$ of the two-level system
and its free evolution frequency $\Delta$ are comparable, $\Gamma\lesssim \Delta$~\cite{Forn-Diaz2017,Magazzu2017}.
A quantum simulation of this region with superconducting microwave circuits
can be done by connecting a superconducting qubit to an open transmission line~\cite{KondoMicrowaves1,KondoMicrowaves2,Forn-Diaz2017,Magazzu2017}.
Very strong couplings (combined with high qubit anharmonicities) are possible by designing system characteristic impedances
comparable to the resistance quantum $R_{\rm Q}=h/(2e)^2$~\cite{KondoMicrowaves1,KondoMicrowaves2,Forn-Diaz2017,Magazzu2017}.
The single Cooper-pair charge $2e$ appears since the anharmonicity of the system is ultimately based on Cooper-pair tunneling across a Josephson junction. For the two-level approximation to hold even under strong dissipation, Cooper-pair tunneling must remain the dominant mechanism. In other words, the coupling strength must be smaller than the qubit anharmonicity, such that only very non-linear qubits such as flux-based qubits are compatible with reaching the ultra-strong coupling regime in the laboratory frame.

Besides increasing the coupling strength via sample design, it also can be effectively increased
by creating a Hamiltonian in the rotating frame, based on the application of Rabi drives~\cite{Ballester2012,Mezzacapo2014}.
In the effective frame, the sub-system energies of the original problem are down-converted to lower frequencies,
while the coupling strength is preserved up to a factor of two.
Applying this approach, an effective ultra-strong coupling between a microwave resonator and a superconducting qubit
has been demonstrated recently also experimentally~\cite{Jochen2017,Langford2017}.
Here, the original qubit-resonator system in the lab frame needs to be only in the strong-coupling regime.
In this work, we study an extension of this approach to a continuous-mode environment, yielding the spin-boson model. Recently, related approaches to effectively achieve ultra-strong coupling have been proposed based on parametric driving~\cite{Qin2018,Leroux2018}.

In this article, we study theoretically a realization of the spin-boson model with strong system-environment couplings
using a superconducting qubit coupled to an engineered environment of bosonic modes.
In analogy to the approach described in Refs.~\cite{Ballester2012,Jochen2017},
we propose to construct an effective spin-boson Hamiltonian in the rotating frame.
The bosonic environment is realized via a set of individual microwave resonators that reside in a restricted frequency range.
We discuss in detail how the microwave circuit maps onto the spin-boson model discussed in literature.
While in principle any bosonic environment can be engineered with the proposed method,
we consider the construction of an environment with an ohmic spectral function that allows for probing localization dynamics of the spin-boson model.
We find that the localization regime
appears at strong coupling between the qubit and individual bosonic modes, which is experimentally feasible to achieve.

We also discuss how the resulting spin dynamics can be probed by standard readout techniques from circuit QED.
In particular, the down-conversion of system frequencies allows for tracking the spin-relaxation dynamics in real time.
We also discuss an experimental implementation, where the bosonic environment and the qubit are fabricated on two separate chips in a modular approach.  This setup allows for probing the system more rigorously, by characterizing both the qubit and the environmental properties in separate experiments.

The article is organized as follows: In Sec.~\ref{sec:UltraStrongCoupling},
we introduce the spin-boson problem in the notation widely used in literature, and how it maps to the notation and methods used in this article.
We 
briefly go through central results and predictions of the spin-boson model.
In particular, in Sec.~\ref{sec:EffectiveHamiltonian}, we show how the effective spin-boson coupling strength can be tailored by two-tone driving.
In Sec.~\ref{sec:CircuitModel}, we introduce an implementation of the spin-boson model by a superconducting transmon qubit coupled to a microwave circuit. We show how the impedance of the environment is related to the spectral density in the spin-boson model
and discuss in detail how the impedance affects to transmon. 
In Sec.~\ref{sec:ActualDesign}, we analyze how a set of microwave resonators can be used to tailor an ohmic spectral density
in the rotating frame with Kondo parameter $\alpha\sim 1$.
In Sec.~\ref{sec:Experiment}, we
provide a description of an experimental realization based on a modular flip-chip approach
and introduce measurement pulse sequences that can be used to probe spin dynamics with different initial conditions.
Conclusions and discussion are given in Sec.~\ref{sec:Conclusions}.

\section{Spin-boson model}\label{sec:UltraStrongCoupling}
We start our analysis by introducing the Hamiltonian and the spectral function of the spin-boson problem.
After this, in Sec.~\ref{sec:ResultsForDifferentEnvironments}, we go through central results and predictions of spin-boson problem obtained in literature~\cite{Spin_Boson_Rev} and discuss what are the corresponding quantities to be measured in our realization.
In Sec.~\ref{sec:EffectiveHamiltonian}, by applying the method described in Refs.~\cite{Ballester2012,Jochen2017}, we derive an effective spin-boson Hamiltonian in the rotating frame with decreased sub-system energies. Finally, in Sec.~\ref{sec:ErrorEstimation}, we analyze the limits of validity of the given derivation.

\subsection{Spin-boson Hamiltonian and the spectral density}
Here, we introduce the spin-boson Hamiltonian in the notation as widely used in earlier literature.
After this we discuss how it maps to the notation used in this article.
The notation and methods used throughout the remainder of this article matches to the standard one used in
superconducting microwave circuits and therefore more directly allows us to relate properties of the spin-boson model to
proposed experimental realization.

\subsubsection{Notation in literature}\label{sec:NotationLiterature}
In earlier literature, the spin-boson model is often introduced by starting from the Hamiltonian~\cite{Spin_Boson_Rev,Book_Weiss}
\begin{eqnarray}\label{eq:SpinBosonLiterature}
\hat H_{\rm SB}&=&-\frac{\hbar\Delta}{2} \hat\sigma_x+\frac{\epsilon}{2} \hat\sigma_z + \frac{q_0}{2}\hat \sigma_z \sum_i  c_i \hat x_i + \hat H_{\rm bath}     \\
\hat H_{\rm bath}&=&\sum_i \left[ \frac{1}{2}m_i \omega_i^2\hat x_i^2+\frac{1}{2m_i}\hat p_i^2  \right]  \, .
\end{eqnarray}
The two-level system, described by the Pauli matrices $\hat \sigma_i$, 
may be regarded as two trapped positions of a virtual particle in a certain potential landscape. The variable $q_0$ denotes a trapping distance, $\Delta$ a hopping rate, and $\epsilon$ characterizes the energy difference. The environment perceives the location of the particle and thereby couples to $\hat\sigma_z$. The free evolution of the environmental coordinate operators $\hat x_i$ is defined by the quadratic harmonic oscillator Hamiltonian $\hat H_{\rm bath}$.

A central function of the theory is the spectral density of the environment, defined formally as
\begin{eqnarray}\label{eq:SpectralDensity}
J(\omega)=\frac{\pi}{2}\sum_i\frac{c^2_i}{m_i\omega_i}\delta(\omega-\omega_i)\, .
\end{eqnarray}
The spectral function $S(\omega)$  of the collective bath operator,
\begin{eqnarray}\label{eq:CouplingOperator}
\hat X = \sum_i c_i\hat x \, ,
\end{eqnarray}
is a function of temperature $T$ and $J(\omega)$, and reads
\begin{eqnarray}\label{eq:CorrelatorX}
S(\omega)=\left\langle  \hat X(t) \hat X(0) \right\rangle_\omega &=& \frac{2\hbar J(\omega)}{1-\exp\left( -\frac{\hbar\omega}{k_{\rm B}T} \right)} \, .
\end{eqnarray}
Together with the parameter $q_0$, see Eq.~(\ref{eq:SpinBosonLiterature}), 
the spectral function includes all relevant information of the effect of the environment on the two-level system. 
The fundamental reason is that the environmental fluctuations satisfy Gaussian statistics. Accordingly,
the Wick's theorem is valid and the time evolution of the reduced density matrix of the two-level system is fully described by two-time correlation functions of the environmental coupling operator.

\subsubsection{Notation in this article}\label{sec:NotationHere}
When superconducting qubits are capacitively or inductively coupled to microwave cavities, their dipole moment couples to the electric or magnetic field of the cavity. Since the dipole coupling is considered transversal, it is intuitive to write the coupling term proportional to a $\hat \sigma _x$ operator. Therefore, even though circuit QED systems consisting of a superconducting qubit coupled to a set of microwave resonators are described by the spin-boson Hamiltonian in Eq.~(\ref{eq:SpinBosonLiterature}), their Hamiltonian is usually written down in a notation where the definition of $\hat\sigma_x$ and $\hat\sigma_z$ are interchanged,
most typically in the context of the Jaynes-Cummings model~\cite{Wallraff2008}. In the case of a transmon qubit~\cite{Transmon}, the two energy levels correspond to two eigenstates of a virtual particle in the same potential minimum.

To keep the notation comparable with Sec.~\ref{sec:NotationLiterature}, we define the system parameters analogously as above.
We then consider establishing the spin-boson Hamiltonian
using a superconducting qubit with  energy splitting $\hbar\Delta$, coupled to a set of microwave resonators, described by the total Hamiltonian
\begin{eqnarray}\label{eq:SpinBosonOur}
\hat H=\frac{\hbar\Delta}{2} \hat\sigma_z + \frac{q_0}{2}\hat \sigma_x \sum_i g_i \left( \hat b_i +\hat b_i^\dagger\right)  + \sum_i \hbar\omega_i\hat b^\dagger_i\hat b_i   \, .
\end{eqnarray}
This corresponds to the case $\epsilon=0$, which is the regime that shows the physically most relevant and non-trivial behavior~\cite{Spin_Boson_Rev}. This Hamiltonian is well implemented by a quantum circuit based on the transmon qubit~\cite{Transmon}. 
The spectral density, defined in Eq.~(\ref{eq:SpectralDensity}), becomes
\begin{eqnarray}\label{eq:SpectralDensityHere}
J(\omega)=\frac{\pi}{\hbar}\sum_i g^2_i \delta(\omega-\omega_i )\, .
\end{eqnarray}

We note that
the coupling parameter $q_0$ could also be incorporated in the definition
of the coupling strengths $g_i$. Our separation is meaningful only when the variables $c_i \hat x_i= g_i \left( \hat b_i +\hat b_i^\dagger\right)$ correspond to certain physical quantities.
In this article, we fix the bath coordinates $\hat x_i$ to correspond to voltage fluctuations across the two capacitors of the qubit,
\begin{eqnarray}\label{eq:Identification}
\hat X(t) \equiv \hat V(t) \, .
\end{eqnarray}
Therefore, $q_0$ has the dimension of charge.
It describes an effective charge shift of the artificial atom between its two states as seen by the environment.
The variable $q_0$ then absorbs all the information of the qubit and how it couples to the voltage fluctuations:
the following results are thereby valid, in principle, for arbitrary superconducting qubits with appropriate adaptations of 
coupling parameter $q_0$. 
Within this identification we then write,
\begin{eqnarray}\label{eq:FluctuationsSpinBoson}
\left\langle  \hat V(t) \hat V(0)  \right\rangle_\omega  &=&\frac{2\hbar J(\omega)}{1-\exp\left( -\frac{\hbar\omega}{k_{\rm B}T} \right)}
\end{eqnarray}
following from Eq.~(\ref{eq:CorrelatorX}) and the identification made in Eq.~(\ref{eq:Identification}).

\subsection{Different bath spectral functions and predictions for the relaxation dynamics of the spin-boson model}\label{sec:ResultsForDifferentEnvironments}
\begin{figure}
\includegraphics[width=\columnwidth]{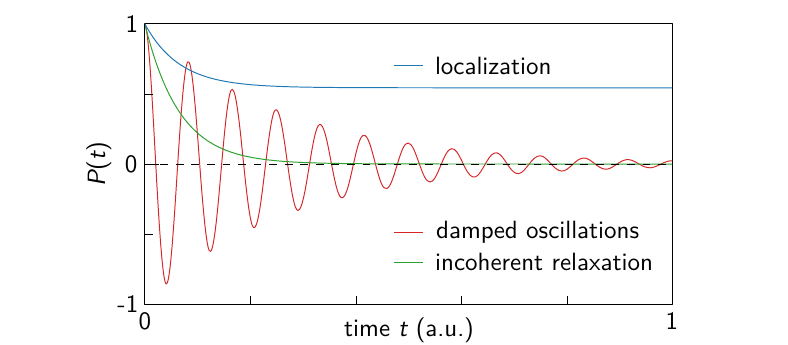}
\caption{Qualitative behavior of spin dynamics in the three main regimes of the spin-boson model with an ohmic environment ($s=1$).
The probability $P(t)$ corresponds in the proposed system to the expectation value $P(t)=\langle \hat \sigma_x(t)\rangle$, when 
initialized to the $+1$ eigenstate of $ \sigma_x$ at $t=0$.
For $\alpha<0.5$,  (damped) oscillations prevail when $\hbar\Delta_{\rm rn}\gtrsim\alpha k_{\rm B}T$, but change to incoherent relaxation
when $\hbar\Delta_{\rm rn}\lesssim\alpha k_{\rm B}T$ (exponential decay to zero).
Localization effect leads to a decay of $P(t)$ towards a finite value and occurs for $\alpha\geq 1$ and $T=0$. In other regimes, the system exhibits incoherent relaxation with subtle forms of the decay rate~\cite{Spin_Boson_Rev}.}\label{fig:OhmicRegimes}
\end{figure}

In the following, we briefly go through some central predictions made for the spin dynamics when interacting with bosonic environments of different spectral functions. We explain how these predictions correspond to dynamics in the considered circuit QED system. A more detailed explanation of an experimental realization is given in Sec.~\ref{sec:Experiment}.
Central predictions for an ohmic environment are summarized qualitatively in Fig.~\ref{fig:OhmicRegimes}.

\subsubsection{Measured quantities}
In the spin-boson model, a widely studied effect is the hopping dynamics between 
the two trapped positions of the fictitious particle (connected by the hopping amplitude $\Delta$) under a perturbation caused by coupling to the environment.
Here, we are not interested in the environment itself, but in the short and intermediate time-scale evolution of the system
when subjected to a certain initial condition. 
The long-time behavior is also interesting to study 
but can be much more challenging to observe in experiment.
The theoretical restrictions to small and intermediate time scales practically
correspond to the experimental restrictions due to the finite initialization time
and finite decoherence time of the superconducting qubit, correspondingly.

We consider now the notation introduced in Sec.~\ref{sec:NotationHere} and follow the discussion given in Ref.~\cite{Spin_Boson_Rev}.
If inserted initially in the left-hand side well, the probability the particle to be found from this well again at some later time
depends on the hopping amplitude and interaction with the environment.
(For a rigorous mathematical definition of the problem, particularly the initialization of the system, see Ref.~\cite{Spin_Boson_Rev}.)
Such population dynamics 
corresponds in our notation to the initialization of the system at $t=0$ to an eigenstate of operator $\hat\sigma_x$ 
 and measuring the value of $\hat \sigma_x$ at certain later time $t>0$,
\begin{eqnarray}
P(t)=\langle\hat\sigma_x(t)\rangle \, .
\end{eqnarray}
Ideally, in the absence of interaction, we get (defining the left-hand side as $+1$ eigenstate of $\hat\sigma_x$)
\begin{eqnarray}
P(t)=\cos\Delta t \, .
\end{eqnarray}
When interacting with the environment, the hopping can become damped, over-damped, or even totally forbidden (localization).

We note that in our realization, we are naturally not restricted to the theoretical scenarios in the literature: one can probe both $\hat\sigma_z$ and $\hat\sigma_x$ with different initialization conditions, for the two-level system as well as for the bath (see Sec.~\ref{sec:Experiment}).
The exact initialization of the bath affects the results essentially in the case of strong couplings, while it is not a requirement for the
 observation of the following effects (in particular the localization).

 

\subsubsection{Relaxation dynamics for different environments}\label{sec:RelaxationDynamics}
A central example of the spin-boson model is the ohmic environment, which is described by a linear spectral density
\begin{eqnarray}
J(\omega) &=& \eta \omega F_{\rm c}(\omega)\, .
\end{eqnarray}
Here we have introduced a cut-off function $F_{\rm c}(\omega)$.
For instance, this can be an exponential drop $F_{\rm c}(\omega)=e^{-\omega/\omega_{\rm c}}$ or a sharp cut-off $F_{\rm c}(\omega)=\Theta(\omega_{\rm c}-\omega)$, with cut-off frequency $\omega_{\rm c}$, respectively.
An important parameter describing the coupling between the system and the environment in the ohmic case is the Kondo parameter
\begin{eqnarray}
\alpha &=& \eta \frac{q_0^2}{2\pi\hbar}\, .
\end{eqnarray}
It can be qualitatively interpreted as an environment-induced decay rate $\Gamma$ normalized by the internal precession frequency $\Delta$, $\alpha\sim \Gamma/\Delta$:
In the limit $\alpha\ll 1$, it directly corresponds to an inverse quality factor of the weakly perturbed two-level system, as derived in Sec.~\ref{sec:KondoParamter}, and
a similar result can also hold for the quantum two-level system with $\alpha\sim 1$~\cite{Forn-Diaz2017,Magazzu2017},
even though here a separation between the system and environment dynamics is not necessary that clearly defined.

It has been understood that
we have practically two independent variables that define the solution of the problem: the interaction strength $\alpha$ and the (bath renormalized)
two-level system energy $\Delta_{\rm rn}$~\cite{Spin_Boson_Rev}.
Under the influence of the environment, many qualitatively different behaviors of the well-hopping dynamics can occur.
For $\alpha<1/2$ we can have damped oscillations ($\Delta_{\rm rn }\gtrsim k_{\rm B}T\alpha$) changing to incoherent relaxation ($\Delta_{\rm rn }\lesssim k_{\rm B}T\alpha$).
For $\alpha>1/2$, all dynamics are expected to be incoherent.
In the regime $\alpha\geq 1$ and $T=0$, one expects a total suppression of hopping, whereas for $T\gtrsim 0$ very slow thermal relaxation should occur~\cite{Spin_Boson_Rev}.
In the simple ohmic case with a linear increase of $J(\omega)$, we therefore expect very different types of behavior in various parameter regimes.
The regimes are summarized in Fig.~\ref{fig:OhmicRegimes}.

It can be helpful to mention that
the localization mechanism in the spin-boson model 
is closely related to Coulomb blockade effect in superconducting tunnel junctions, i.e.,
Cooper-pair tunneling across a Josephson junction that is voltage-biased in series with an electromagnetic environment.
When the environmental (zero-frequency or characteristic resonator) impedance is comparable with the resistance quantum $R_{\rm Q}=h/4e^2$, the system enters the Coulomb blockade regime, where charge tunneling is strongly suppressed or even completely prohibited~\cite{Schon1990,Ingold1992,Dambach2015,Leppakangas2016}.

The model for general power-law behavior of $J(\omega)$ is conveniently written in the form
\begin{eqnarray}
J(\omega) &=& A_s \omega^s \omega_{\rm c}^{1-s} F_{\rm c}(\omega)\, .
\end{eqnarray}
Here, the case $s < 1$ is called the sub-ohmic regime and $s > 1$ is referred to as the super-ohmic regime.
In particular, $s=0$ with $T>0$ has been used as a model for $1/f$ noise~\cite{Shnirman2002,Marthaler2016}. The super-ohmic case
appears in the electron tunneling in solids with coupling to a (three-dimensional) phononic bath.
The extra scaling factor $\omega_{\rm c}^{1-s}$ has been introduced
so that we can define a dimensionless variable ${\cal A}=A_sq_0^2/2\pi\hbar$, in analogy to the Kondo parameter $\alpha$.
However, it has less physical meaning here as in the ohmic case. 
It also always appears together with the introduced scaling by the cut-off,
${\cal A}\omega_{\rm c}^{1-s}$~\cite{Spin_Boson_Rev}.
In a rough overall picture, the super-ohmic case shows mostly damped oscillations and does not exhibit localization,
whereas the sub-ohmic case is less trivial: it is localized for weak tunneling amplitudes $\Delta$ (depending on ${\cal A}$)
but even there, in non-equilibrium, can show coherent oscillations~\cite{Bulla2007}. Also as opposed to the ohmic case, here exists
more than one relevant energy scale of coherent dynamics.


\subsection{Simulation in the rotating frame}\label{sec:EffectiveHamiltonian}
Here, we show how to establish an effective spin-boson Hamiltonian in the rotating frame by additional microwave driving.
We take use of modified interaction during driven evolution of the two-level system~\cite{Ithier2005}.
An important detail of the following derivation is that even though rotating-wave approximations (RWA)
can be taken in various places of the derivation, it cannot be taken for the final effective Hamiltonian,
where the effect of counter-rotating terms can be essential.


\subsubsection{Two-tone driving}
Following Refs.~\cite{Ballester2012,Jochen2017}, we consider driving this system with two Rabi tones, both with transverse coupling to the qubit.
A Hamiltonian that describes such a driven system has the form
\begin{eqnarray}\label{eq:LaboratoryHamiltonian}
\hat H+ \hat H_{\rm d} \, ,
\end{eqnarray}
where the drive is accounted for by the term
\begin{eqnarray}\label{eq:DrivingTerms}
\hat H_{\rm d}=  \hbar \Omega_1 \hat\sigma_x\cos \omega_1 t + \hbar \Omega_2 \hat\sigma_x\cos \omega_2 t \, .
\end{eqnarray}
Here $\Omega_{i}$ is the amplitude and $\omega_{i}$ the frequency of the drive~$i$.
To obtain an immediate feeling of the drive frequencies and amplitudes we use, we note that
in the following scheme we  consider a situation where $\omega_1\gtrsim\omega_2$ and $\Omega_1\gg \Omega_2$.
During the derivation, also the condition $\omega_1-\omega_2=\Omega_1$ is taken to obtain the desired form of the Hamiltonian (see below), and we will have $\omega_i\gg\Omega_i$.
The drive frequency can be assumed to be the qubit frequency in the lab frame, $\omega_1=\Delta$.

We enter now a rotating frame with respect to the stronger transverse drive by performing a unitary transformation according to
\begin{eqnarray}
\hat U=\exp\left[ \mathrm {i}\omega_1 t \left( \sum_i\hat b^\dagger_i \hat b_i +\frac{1}{2}\hat\sigma_z \right) \right]\, .
\end{eqnarray}
This is a combined rotating frame of the two-level system and of all the bosonic modes. The Hamiltonian becomes now
\begin{eqnarray}\label{eq:HamiltonianDerivation}
\frac{\hat H_{1}}{\hbar}&=&\frac{1}{\hbar}\left(\hat U \hat H \hat U^\dagger -\mathrm {i}\hat U \dot{\hat{U}}^\dagger\right) =\frac{\Delta-\omega_1}{2} \hat\sigma_z +\frac{\Omega_1}{2}\hat\sigma_x  \\
&+&   \sum_i (\omega_i-\omega_1)\hat b^\dagger_i\hat b_i + \frac{q_0}{2\hbar}\sum_i g_i\left( \hat b_i \hat\sigma_+  +\hat b_i^\dagger \hat\sigma_- \right)  \nonumber \nonumber \\
&+& \frac{\Omega_2}{2}\left( e^{\mathrm {i}(\omega_1-\omega_2)t}\hat\sigma_+ +e^{-\mathrm {i}(\omega_1-\omega_2)t}\hat\sigma_- \right)  \, . \nonumber
\end{eqnarray}
We have neglected the contributions
\begin{eqnarray}\label{eq:HamiltonianNeglect1}
\hat{O}_1&=& \frac{q_0}{2\hbar}\sum_i g_i\hat\sigma_+ \hat b_i^\dagger e^{2\mathrm {i}\omega_1 t}+\frac{\Omega_1}{2}\hat\sigma_+ e^{2\mathrm {i}\omega_1 t} \\
&+&\frac{\Omega_2}{2}\hat\sigma_+e^{\mathrm {i}(\omega_1+\omega_2)t}+{\rm H.c.} \, . \nonumber
\end{eqnarray}
This can be done if oscillations with the frequencies $2\omega_1$ and $\omega_1+\omega_2$ are much faster than
frequencies $\Omega_1$ and $\Omega_2$.
In addition, coupling to modes in the bosonic bath, with couplings $q_0g_i/\hbar$, is negligible if the bath will include only modes in a small frequency range $\omega_{\rm c}\ll 2\omega_i$.

In Hamiltonian of Eq.~(\ref{eq:HamiltonianDerivation}), the dominant term will be the contribution proportional to $\Omega_1$.
It is then favorable to move to the interaction picture defined by this term. This means performing another unitary transformation,
this time according to
\begin{eqnarray}
\hat U=\exp\left[ \mathrm {i}\frac{\Omega_1}{2} \hat\sigma_x t \right]\, .
\end{eqnarray}
We also choose $\Omega_1=\omega_1-\omega_2$, which leads to 
\begin{eqnarray}\label{eq:EffectiveHamiltonian}
&&\frac{\hat H_{2}}{\hbar}=  \\
&&\frac{\Omega_2}{4} \hat\sigma_z +  \frac{q_0}{2\hbar}\sum_i\frac{g_i}{2} \hat\sigma_x (\hat b^\dagger_i + \hat b_i )  +\sum_i (\omega_i-\omega_1)\hat b^\dagger_i \hat b_i \, . \nonumber
\end{eqnarray}
We have again neglected fast oscillating terms,
\begin{eqnarray}\label{eq:HamiltonianNeglect2}
\hat{O}_2&=& \frac{\Omega_2}{2}\hat\sigma_z\left(\sin^2\Omega_1t+\frac{1}{2}\right)\\
&-&\frac{\Omega_2}{2}\left( \hat\sigma_1 \sin \Omega_1t-\hat\sigma_y\sin 2\Omega_1t ) \right) \nonumber \\
&+&(\Delta-\omega_1)( \hat\sigma_z\cos \Omega_1t +\hat\sigma_y\sin \Omega_1t ) \nonumber \\
&+&\frac{q_0}{4\hbar}\sum_i g_i\left[ \left(\mathrm {i}\hat\sigma_y\cos \Omega_1 t+\mathrm {i}\hat\sigma_z\hat \sin \Omega_1 t \right) b^\dagger_i + {\rm H.c.} \right)  \, , \nonumber
\end{eqnarray}
The first three terms on the right-hand side can be easily dropped with similar assumptions as above.
The implications due to dropping the fourth term need to be analyzed more carefully, done below in Sec.~\ref{sec:ErrorEstimation}.

\subsubsection{Effective Hamiltonian and spectral density}\label{sec:EffectiveSpectralFunction}
We note that the Hamiltonian of Eq.~(\ref{eq:EffectiveHamiltonian})
has the same (non-RWA) interaction term as in Eq.~(\ref{eq:SpinBosonOur}), with modified parameters.
We then have the effective Hamiltonian
\begin{eqnarray}\label{eq:EffectiveHamiltonian2}
\hat H_{\rm eff}&=&\frac{\hbar\Delta^{\rm eff}}{2} \hat\sigma_z \\
&+& \frac{q_0}{2}\hat\sigma_x \sum_i g_i^{\rm eff} \left( \hat b_i +\hat b_i^\dagger\right)  + \sum_i \hbar\omega_i^{\rm eff}\hat b^\dagger_i\hat b_i   \, , \nonumber
\end{eqnarray}
where the new parameters have the form
\begin{eqnarray}
\Delta^{\rm eff}&=&\frac{\Omega_2}{2} \\
\omega^{\rm eff}_i&=&\omega_i-\omega_1 \\
g_i^{\rm eff}&=&\frac{g_i}{2} \, .
\end{eqnarray}
We see that the two-level system and bosonic energies are tunable by the external drives.
Since the coupling  has kept its form (up to a factor of 2), this 
allows for tailoring essentially stronger relative couplings between the system and the environment~\cite{Ballester2012,Jochen2017}.

We also have a new coordinate operator of the environment.
To determine its properties we first write down the solution in the rotating frame
\begin{eqnarray}
\hat V_{\rm eff}(t)&=&\frac{1}{2}\sum_i g_i\left[ \hat b_i e^{-\mathrm{i}(\omega_i-\omega_1) t}+ \hat b_i^\dagger e^{\mathrm{i}(\omega_i-\omega_1) t}\right]  \, . 
\end{eqnarray}
Here the energies $\omega_i-\omega_1$ are the effective energies in the rotating basis, which can  be negative.
The population of these modes 
can be determined from the thermal population in the laboratory frame.
Using the spectral density in the original frame, $J(\omega)$, we get for the thermal average of the correlation function
\begin{eqnarray}
\left\langle \hat V_{\rm eff}(t) \hat V_{\rm eff}(0) \right\rangle_\omega &=& \frac{\hbar}{2}  \frac{ J(\omega+\omega_1) }{1-\exp\left(-\frac{\hbar(\omega+\omega_1)}{k_{\rm B}T}\right)} \, .
\end{eqnarray}
The temperature $T$ is the real temperature of the bath.
In the following, it is safe to assume that the real bath is at the zero temperature
since practically $\omega_1\gg k_{\rm B}T/\hbar$.
We have then
\begin{eqnarray}
\left\langle \hat V_{\rm eff}(t) \hat V_{\rm eff}(0) \right\rangle_\omega &=& \frac{\hbar}{2} J(\omega+\omega_1) \, .
\end{eqnarray}

In order to have an exact connection between the effective system in the rotating frame and the spin-boson model,
the created correlation function in the rotating frame has to simulate a finite temperature bath.
To construct a specific spectral function in the rotating frame with an effective temperature $T_{\rm eff}$, the spectral density in the laboratory frame is required to have a contribution ($\delta\omega>0$) below the frequency of the rotating frame,
\begin{eqnarray}\label{eq:DetailedBalance}
J(\omega_1-\delta\omega) &=&  J(\omega_1+\delta\omega)\frac{1-\exp\left[-\frac{\hbar\delta\omega}{k_{\rm B}T_{\rm eff}}\right]}{\exp\left[\frac{\hbar\delta\omega}{k_{\rm B}T_{\rm eff}}\right]-1}  \, .
\end{eqnarray}
If this is satisfied for certain $T_{\rm eff}$, we have
\begin{eqnarray}
\left\langle \left[ \hat V_{\rm eff}(t), \hat V_{\rm eff}(0)\right]_+\right\rangle_\omega &=& 2\hbar J_{\rm eff}(\delta\omega) \coth \frac{\hbar\delta\omega}{2k_{\rm B}T_{\rm eff}} \, ,
\end{eqnarray}
where we have defined the effective spectral density in the rotating frame
\begin{eqnarray}
J_{\rm eff}(\delta\omega)=\frac{1}{4} J(\omega_1+\delta\omega)\left\{ 1-\exp\left[-\frac{\hbar\delta\omega}{k_{\rm B}T_{\rm eff}}\right]  \right\} \, .
\end{eqnarray}
For $T_{\rm eff}=0$ we have simply 
\begin{eqnarray}
J_{\rm eff}(\delta\omega)=\frac{1}{4} J(\omega_1+\delta\omega) \, .
\end{eqnarray}


We note that even though the connection between these two systems might seem trivial,
just a frequency shift due to the external drive,
it is quite remarkable since it connects two completely different many-body physics problems:
one problem including emission and absorption of photons with same bosonic modes,
and another problem which includes only dissipation to two different set of bosonic modes.
The only property that needs to be satisfied to connect these two problems is the effective detailed balance, Eq.~(\ref{eq:DetailedBalance}).

\subsection{Error estimation}\label{sec:ErrorEstimation}
Here, we sum up the restrictions and the size of errors in the quantum simulation that appear due to the taken approximations
when deriving the effective rotating-frame Hamiltonian.
Errors occur from dropping the terms in Eqs.~(\ref{eq:HamiltonianNeglect1}) and~(\ref{eq:HamiltonianNeglect2}).
Furthermore, errors  also occur due to a finite anharmonicity of the two-level system, which can lead to a finite population of the third level of the
superconducting qubit.

Most terms in Eqs.~(\ref{eq:HamiltonianNeglect1}) and~(\ref{eq:HamiltonianNeglect2}) can be dropped
within the assumptions $\Omega_i/\omega_i\ll 1$ and $(\omega_1-\Delta)/\Omega_1\ll 1$, as well as $\Omega_2/\Omega_1\ll 1$.
These conditions are easily realized in an experiment~\cite{Jochen2017}. However, the most important contribution we neglected was the term
\begin{eqnarray}\label{eq:HamiltonianNeglect3}
\hat{O}= \frac{q_0}{4}\sum_i g_i\left[ \left(\mathrm {i}\hat\sigma_y\cos \Omega_1 t+\mathrm {i}\hat\sigma_z \sin \Omega_1 t \right) b^\dagger_i + {\rm H.c.} \right)  \, .
\end{eqnarray}
This sets a limit to the spectral width and the cut-off of the bath.
This is since the term probes the bath in a completely similar way as the central term
\begin{eqnarray}
\frac{q_0}{4}\sum_i g_i \hat\sigma_x (\hat b^\dagger_i + \hat b_i )\, ,
\end{eqnarray}
in the effective Hamiltonian of Eq.~(\ref{eq:EffectiveHamiltonian2}), but with energies $\Omega_1\pm\Omega_2/2\approx\Omega_1$.

To be more quantitative, let us assume that we have a residue bath density at frequencies close to $\Omega_1$, which we now write in the form
\begin{eqnarray}
J_{\rm eff}(\Omega_1)\approx \frac{2\pi\hbar}{q_0^2}\bar\alpha\frac{\Omega_2}{2} \, .
\end{eqnarray}
The dimensionless variable $\bar\alpha$ then compares the effective qubit frequency $\Omega_2/2$ to the spectral density at frequency $\Omega_1$. 
This gives a bath-induced decoherence rate
\begin{eqnarray}
\bar\Gamma \approx\pi\bar\alpha \frac{\Omega_2}{2} \, .
\end{eqnarray}
In order to have a negligible contribution within the time scale of the effective two-level system oscillations, $1/\Omega_2$, we demand $\bar\alpha\ll 1$.
Similarly, also a finite internal lifetime of the two-level system, due to internal decay mechanisms, limits the simulation
length. Let us denote this rate by $\Gamma_{\rm internal}$. Ideally, we would then like to engineer a bath which does not
limit the decay  and dephasing times of the qubit itself, i.e., we would like to be in the regime
$\bar\Gamma< \Gamma_{\rm internal}\ll \Omega_2/2$.

The second important restriction to the parameter regime
is the finite anharmonicity of the qubit. The anharmonicity is defined as the difference between the first and second energy-level splittings,
\begin{eqnarray}
\hbar\Delta_{\rm an}=\vert(E_2-E_1)-(E_3-E_2)\vert \, .
\end{eqnarray}
Too strong drive can induce transitions to the third state of the artificial atom.
The probability for the artificial atom contributing through the third excited state is roughly
\begin{eqnarray}
P_{\rm error}\sim \left(\frac{\Omega_1}{\Delta_{\rm an}}\right)^2 \, .
\end{eqnarray}
Therefore, a large anharmonicity qubit is favorable in order to avoid a strong additional upper bound in $\Omega_1$.
The qubit anharmonicity depends on the experimental realization. Flux-based qubits can easily reach anharmonicities higher than the lowest energy-level splitting $\Delta_{\rm an}>\Delta$. 
In this article, we consider a realization based on a transmon qubit with $\Delta_{\rm an}\ll \Delta$ for its simple operation without the necessity of biasing~\cite{Transmon}, the feasibility of a straightforward capacitive coupling, and its superior coherence properties.
For a qubit with $ \Delta=2\pi\times 7$~GHz and anharmonicity $ \Delta_{\rm an}=2\pi\times 350$~MHz, a drive with $\Omega_1=2\pi\times 80$~MHz leads to a reasonable low error $P_{\rm error}\sim 0.05$. Combining this with the above analysis,
this would also mean that the bath spectral width has to be smaller than $80$~MHz,
in order
to avoid unwanted transitions due to the term in Eq.~(\ref{eq:HamiltonianNeglect3}).
We would then desire a bath that has a rather sharp cut-off at $\omega_{\rm c}<\Omega_1=2\pi \times 80$~MHz, $F_{\rm c}(\omega)\sim \Theta(\omega_{\rm c}-\omega)$. Later, in Sec.~\ref{sec:ActualDesign}, we show how to build such a bath from a set of microwave resonators.

\section{Implementation of the spin-boson model with a microwave circuit}\label{sec:CircuitModel}

In this section, we study how a superconducting qubit connected to a dissipative microwave-circuit element
can be used to realize the spin-boson Hamiltonian.
We consider explicitly the case of a transmon qubit.
Our main goal is to determine how the parameters of the spin-boson model,
the spectral density $S(\omega)$, the coupling $q_0$, and the qubit energy $\Delta$, depend on the properties of the microwave circuit.
Section~\ref{sec:CircuitAnalysisMainResults} briefly sums up the central results.
In Sec.~\ref{sec:GeneralCircuitHamiltonian}, we describe how to determine the effect of capacitance renormalization
in circuits considered in this article.
In Sec.~\ref{sec:SpinBosonConnectionLab}, we detail the derivation of
the spin-boson parameters $q_0$  and $\Delta$, and in Sec.~\ref{sec:KondoParamter}, we show the derivation of the Kondo parameter $\alpha$.
The approach we use is based on a linear circuit analysis,
but the results can also be derived by an exact Lagrangian quantization~\cite{Yorke1984,Pozar,Wallquist2006,Loudon,Malekakhlagh2016,Parra2017}.
In addition, we provide also a consistency check based on the Born-Markov approach, in Sec.~\ref{sec:GeneralCircuitHamiltonian}.
Even though we explicitly consider a transmon qubit, our formalism is generic and can be extended, in principle, to all superconducting qubit architectures.

\begin{figure}
\includegraphics[width=0.9\columnwidth]{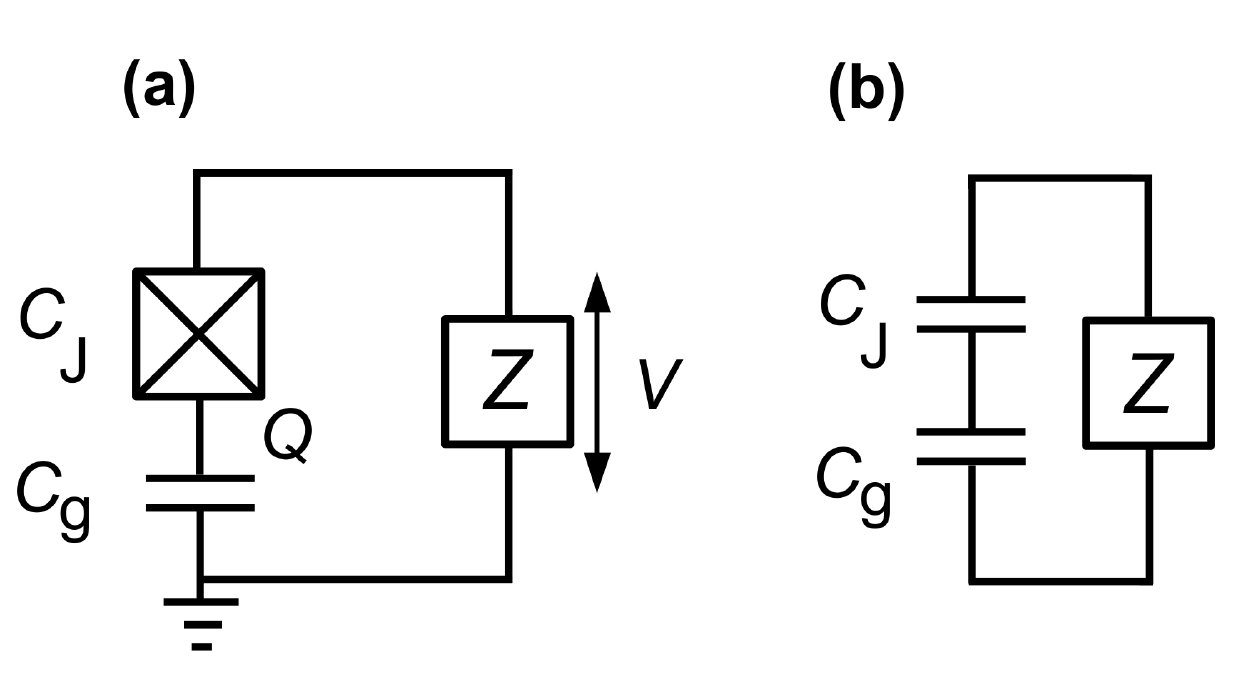}
\caption{(a) A model of a transmon qubit connected to an impedance $Z(\omega)$. The charge $Q$ on the island between the Josephson junction (crossed box) and the ground capacitor $C_{\rm g}$ is a conjugated variable to the phase across the Josephson junction, providing
anharmonic energy levels and an effective two-level system.
The impedance $Z$ induces voltage fluctuations ($V$) and dissipation. (b) The circuit that defines the spectral density,
Eqs.~(\ref{eq:InteractionCapacitance}-\ref{eq:VoltageCorrelatorCircuit2})} \label{fig:Transmon}
\end{figure}

\subsection{Spectral density and the system-bath interaction}\label{sec:CircuitAnalysisMainResults}
Our superconducting qubit couples to environmental voltage fluctuations $ \hat V(t)$, that causes dissipation. The quantity that describes its effect is the spectral density
$S(\omega)=\left\langle \hat V(t)\hat V(0) \right\rangle_\omega$.
There are several equivalent ways of determining this quantity for microwave circuits,
which basically all seek for the eigenmodes of the relevant (non-interacting) linear system.
In this article, we assume that we know the impedance $Z(\omega)$ of the linear circuit connected to the superconducting qubit,
an example being the circuit we consider in Sec.~\ref{sec:ActualDesign}.
Guidelines for a determination of the spectral density in open circuits is given in Appendix~A as well as in other Refs.~\cite{Yorke1984,Wallquist2006,Loudon,Pozar,Malekakhlagh2016,Parra2017}

Voltage fluctuations across the impedance are described by the operator $\hat V$.
The exact circuit diagram of the considered setup is shown in Fig.~\ref{fig:Transmon}(a).
Generally, voltage fluctuations in a linear (free-evolution) electric circuit satisfy the quantum fluctuation-dissipation theorem~\cite{Ingold1992},
\begin{eqnarray}\label{eq:VoltageCorrelatorCircuit}
\left\langle \hat V(t)\hat V(0) \right\rangle_\omega=\frac{2\hbar\omega {\rm Re}[Z_{\rm eff}(\omega)]}{1-e^{-\beta\hbar\omega}} \, .
\end{eqnarray}
In this free evolution solution, where the transmon island charge is set to zero (see below),
the impedance $Z(\omega)$ sees a parallel capacitance  $C_{\rm int}$, which is the effective qubit capacitance~\cite{Ingold1992,Shnirman2003,Leppakangas2008},
\begin{eqnarray}\label{eq:InteractionCapacitance}
C_{\rm int}=\left(C_{\rm J}^{-1}+C_{\rm g}^{-1}\right)^{-1} \, .
\end{eqnarray}
$C_{\rm J}$ and $C_{\rm g}$ denote the capacitances of the Josephson junction and the capacitance to ground, respectively. The effective impedance of the environment, to be used in Eq.~(\ref{eq:VoltageCorrelatorCircuit}), assumes the form
\begin{eqnarray}\label{eq:ImpedanceEffective}
Z_{\rm eff}^{-1}(\omega)=\mathrm{i}\omega C_{\rm int}+Z^{-1}(\omega) \, .
\end{eqnarray}
The equivalent circuit is shown in Fig.~\ref{fig:Transmon}(b). Note that the inductance of the Josephson junction, which determines the qubit dynamics, does not enter the calculation of $Z_{\rm eff}^{-1}(\omega)$, but only the effective qubit capacitance $C_{\rm int}$ that shunts the effective bath impedance.

We also note that the scenario where a bath circuit is used to tailor a dissipative qubit environment is fundamentally different from the case where a certain impedance is used to filter microwave transmission. 
The reason is a different boundary condition at the qubit: In the case of the tailored bosonic environment, radiation reflects at the capacitor $C_{\rm int}$, whereas in the case of a microwave filter, we would have an impedance-matched load and no reflection.

A direct comparison of Eqs.~\eqref{eq:FluctuationsSpinBoson},~\eqref{eq:VoltageCorrelatorCircuit} yields the relation between the spectral density of the spin-boson model and the effective impedance,
\begin{equation}\label{eq:VoltageCorrelatorCircuit2}
J(\omega)=\omega{\rm Re}[Z_{\rm eff}(\omega)] \, .
\end{equation}
This equation is central for experimentally tailoring a bosonic environment, relating the effective impedance to the resulting spectral density $J(\omega)$.
It has also been shown recently that the parallel contribution $C_{\rm int}$ in the spectral density
is indeed an essential quantity for a consistent description of such systems in all parameter regimes~\cite{Malekakhlagh2016,Parra2017}.

In the considered circuit,
the transmon interacts with the environmental voltage fluctuations through the operator~\cite{Shnirman2003,Leppakangas2008}
\begin{eqnarray}
\hat H_{\rm int}&=&\beta\hat Q\hat V\equiv\hat Q_{\rm int}\hat V\label{eq:InteractionCharge} \\
\beta &=&\frac{C_{\rm g}}{C_{\rm J}+C_{\rm g}}  \, .
\end{eqnarray}
Here $\hat Q$ is the charge operator of the transmon island.
The interaction charge, $\hat Q_{\rm int}$, accounts for an internal transmon-qubit capacitive shunting through parameter $\beta$,
reducing the coupling to the island charge $\hat Q$~\cite{Transmon}. The parameter $\beta$ is not affected by renormalization effects.
However, for determination of the resulting spin-boson Hamiltonian parameter $q_0$, one generally needs to consider also the possible
qubit-capacitance renormalization due to coupling to the impedance, as analyzed in Sec.~\ref{sec:GeneralCircuitHamiltonian}.
The final result reads
\begin{eqnarray}\label{eq:ConnectionqPreliminary}
q_0= 2e\beta \sqrt{\frac{R_{\rm Q}}{\pi Z_{\rm J}}} \, .
\end{eqnarray}
Here, the characteristic impedance of the transmon is defined as $Z_{\rm J}=R_{\rm Q}\sqrt{2E_C/\pi^2E_{\rm J}}$, where
$E_{\rm J}$ is the Josephson coupling energy, $E_C=e^2/2(C_{\rm J}+C_{\rm g}^0)$
the charging energy, and the effective ground capacitance $C_{\rm g}^0$
depends on the realization (see Sec.~\ref{sec:GeneralCircuitHamiltonian}). In the simplest case $C_{\rm g}^0= C_{\rm g}$.
Finally, the normalized two-level system energy $\Delta$ for typical transmon parameters becomes~\cite{Transmon}
\begin{eqnarray}
\Delta &\approx& \frac{1}{\hbar}\sqrt{8E_{\rm J}E_C} \, .
\end{eqnarray}
In the following section, we show how to determine $E_C$ and demonstrate that the given identifications are consistent with
the alternative approach of including the interaction term of Eq.~(\ref{eq:InteractionCharge}) using a Born-Markov approximation.
It is also consistent with the exact derivation when using an open-circuit, given in Appendix~A.

\subsection{Capacitance renormalization}\label{sec:GeneralCircuitHamiltonian}
The impedance $Z(\omega)$ can affect to the Hamiltonian of the transmon.
The effect is generally twofold: it renormalizes (i)  the effective transmon capacitance 
and (ii) the Josephson coupling energy $E_{\rm J}$.
The  effect (i) is analogous to mass renormalization in the spin-boson model~\cite{Spin_Boson_Rev} and can be here significant.
The effect (ii) is analogous to tunneling-amplitude renormalization in the spin-boson model, before going into the spin-boson representation~\cite{Spin_Boson_Rev},
and stays here small due to considered small environmental impedances, $Z\ll R_{\rm Q}$, and low qubit energies in comparison to the superconducting energy gap.

\subsubsection{Hamiltonian of an isolated transmon}\label{sec:HamiltonianOfTransmon}
The Hamiltonian of a superconducting artificial atom can be derived by applying a Lagrangian formalism to electric circuits~\cite{Devoret}.
The Hamiltonian of an isolated transmon is of the form~\cite{Transmon}
\begin{eqnarray}\label{eq:TransmonHamiltonian}
\hat H^{\rm isolated}_{\rm tr}=-E_{\rm J}\cos \hat \varphi + \frac{\hat Q^2}{2(C_{\rm J}+C_{\rm g})}     \, ,
\end{eqnarray}
The first term on the right-hand side describes Cooper-pair tunneling across the superconducting junction
as a function of the superconducting phase difference $\hat \varphi$ across the Josephson junction.
The second term
describes the capacitive (Coulomb) energy related to the island charge $Q$.
In this isolated circuit, the effective island capacitance is the sum of $C_{\rm J}$ and $C_{\rm g}$.
The phase and the charge are conjugated variables,
\begin{eqnarray}
\left[\frac{\hat{Q}}{2e},e^{\mathrm{i}\hat{\varphi}} \right] =e^{ \mathrm{i}\hat{\varphi}} \, .
\end{eqnarray}
The commutation relation is presented in this (periodic) form since the island charge takes only values that are multiples of $2e$,
or equivalently, the phase distribution is here by definition $2\pi$-periodic.

\subsubsection{Accounting for the counter-term}\label{sec:GeneralCircuitHamiltonianMethod}
Finding the capacitance renormalization is analogous to identifying the 'counter-term' in general system-reservoir models~\cite{Spin_Boson_Rev,Book_Weiss}.
In this analysis, we study two equivalent forms of the total Hamiltonian,
\begin{eqnarray}
\hat H_{\rm total} &=&\hat H_{\rm tr}+\hat H_{\rm bath}+ \hat H_{\rm int}\label{eq:TotalHamiltonianForm1}  \\
\hat H_{\rm total} &=&\hat H_{\rm tr}^0+\hat H_{\rm bath}+\left[\hat H_{\rm int} + \hat H_{\rm ct}\right] \, , \label{eq:TotalHamiltonianForm2}
\end{eqnarray}
where then
\begin{eqnarray}\label{eq:RenormalizationFormal}
\hat H_{\rm tr}^0 =\hat H_{\rm tr}-\hat H_{\rm ct}  \, .
\end{eqnarray}
In addition to the qubit, bath, and interaction Hamiltonians, we have introduced a term $\hat H_{\rm ct}$,
counteracting to the qubit Hamiltonian renormalization (coherent embedding of the environment) coming from the interaction term $\hat H_{\rm int}$.
{\em It is here the interaction-normalized Hamiltonian $\hat H_{\rm tr}^0$ that should be used when theoretically reducing the transmon to a two-level system
and whose dynamics is observed in the experiment.}


Strictly speaking, the renormalization is determined theoretically by first evaluating Hamiltonian of Eq.~(\ref{eq:TotalHamiltonianForm1}),
for example, by using a Lagrangian approach (Appendix A),
and then estimating the embedding due to the interaction term $\hat H_{\rm int}=\hat Q_{\rm int}\hat V$.
However, we find that in circuits we consider the contributions $\hat H_{\rm ct}$, $\hat H_{\rm tr}$ and $\hat H_{\rm tr}^0$
can be deduced more straightforwardly from the following coherent solutions:

\begin{itemize}

\item  The solution when the resistivity is put to zero, giving $H_{\rm tr}^0$.

\item  The solution when the resistive part is disentangled from the circuit, for example, with an additional
capacitor $C_{\rm dis}\rightarrow 0$ in series with the resistor, giving $H_{\rm tr}$.

\end{itemize}


\begin{figure}
\includegraphics[width=0.6\columnwidth]{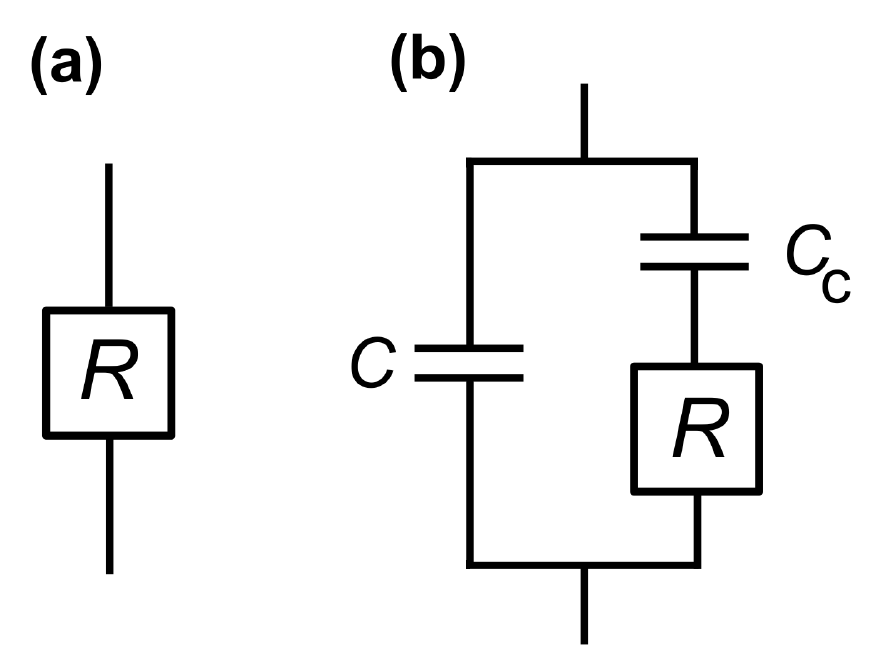}
\caption{Two environmental impedances $Z(\omega)$, whose capacitance renormalization is considered explicitly in this section.} \label{fig:SimpleImpedances}
\end{figure}

To illustrate the mathematics of this approach, let us consider the simple case of a bare ohmic impedance $Z(\omega)=R$.
We then first identify
the Hamiltonian of the circuit when resistivity is set to zero. This fully coherent system corresponds to the one in Eq.~(\ref{eq:TransmonHamiltonian}),
\begin{eqnarray}
\hat H_{\rm tr}^0=\hat H^{\rm isolated}_{\rm tr}
\end{eqnarray}
In the second stage,
we identify the transmon Hamiltonian when disconnected from the resistor lead, which has the form (Appendix A)
\begin{eqnarray}\label{eq:TransmonHamiltonian2}
\hat H_{\rm tr}=-E_{\rm J}\cos \hat \varphi + \frac{\hat Q^2}{2C_{\rm J}}     \, .
\end{eqnarray}
Using this we then find for the difference
\begin{eqnarray}\label{eq:TransmonRenormalization}
\hat H_{\rm ct} &=& \hat H_{\rm tr}- \hat H_{\rm tr}^0  \nonumber   \\
&=&  \frac{\hat Q^2}{2C_{\rm J}}-\frac{\hat Q^2}{2(C_{\rm J}+C_{\rm g})} = \frac{\hat Q_{\rm int}^2}{2C_{\rm int}} \, . 
\end{eqnarray}

Let us then consider the circuit shown in
Fig.~\ref{fig:SimpleImpedances}(b), which is analogous to our proposal presented in Sec.~\ref{sec:ActualDesign}.
When the resistance is put to zero, an environmental
capacitive remains with contribution $C+C_{\rm c}$, leading to
\begin{eqnarray}
\hat H_{\rm tr}^0=-E_{\rm J}\cos \hat \varphi  + \frac{\hat Q^2}{2(C_{\rm J}+C_{\rm g}^0)}     \, , \label{eq:TransmonHamiltonianCorr2}
\end{eqnarray}
where
\begin{eqnarray}\label{eq:TransmonHamiltonianCorr22}
C_{\rm g}^0=\left( C_g^{-1} + (C+C_{\rm c})^{-1} \right)^{-1} \, .
\end{eqnarray}
In the second stage, we get for the Hamiltonian corresponding to the disconnected resistor
\begin{eqnarray}
\hat H_{\rm tr}=-E_{\rm J}\cos \hat \varphi  + \frac{\hat Q^2}{2(C_{\rm J}+C_{\rm g'})}     \, ,
\end{eqnarray}
where we have defined an effective gate capacitance
\begin{eqnarray}
C_{\rm g '}=\left( C_g^{-1} + C^{-1} \right)^{-1} \, .
\end{eqnarray}
This is since $C$ appears in series connection with $C_g$. 
To evaluate the counter-term, let us consider explicitly the case $C\rightarrow 0$. (The analysis of this section also holds also for $C\neq 0$.)
We get for the difference
\begin{eqnarray}\label{eq:RNFigre1ub}
\hat H_{\rm ct} &=& \hat H_{\rm tr}-\hat H_{\rm tr}^0  \nonumber   \\
&=& \frac{\hat Q^2}{2(C_{\rm J}+C_{\rm g'})}-\frac{\hat Q^2}{2(C_{\rm J}+C_{\rm g}^0)} \nonumber\\
&=&   \frac{C_{\rm c}}{C_{\rm c}+C_{\rm int}}\frac{1}{2C_{\rm int}} \hat Q^2_{\rm int}  \, . 
\end{eqnarray}

To show that the above results are sound, we can estimate
the embedding due to the interaction term $\hat H_{\rm int}=\hat Q_{\rm int}\hat V$ by an alternative method, using a Born-Markov master equation.
Such an approach assumes that the effect of the environment (beyond the counter-term) is weak,
but its result is valid also more generally since the embedding of the environment is the same for all $R$.
Here we start from the Hamiltonian $\hat H_{\rm tr}$, where the resistor lead is decoupled from the transmon,
and estimate the renormalization explicitly. Considering the circuit of Fig.~\ref{fig:SimpleImpedances}(a), we then use
the property that for an ohmic environment with resistance $R$ and cut-off defined by the parallel capacitor,
$(RC_{\mathrm{int}})^{-1}=\omega_{\rm c}$, 
we have transition rates and energy-level renormalization terms
\begin{eqnarray}
&&\lim_{s\rightarrow 0} \int_0^{\infty} dt e^{ \mathrm{i} (\omega  + \mathrm{i} s)t}\left\langle \hat V(t) \hat V(0)\right\rangle = \frac{\hbar\omega}{1-e^{-\beta\hbar\omega}}{\rm Re}[Z(\omega)] \nonumber \\
&-& \mathrm{i} \frac{\hbar\omega_{\rm c}}{2}{\rm Re}[Z(\omega)]+ \mathrm{i} \frac{\hbar\omega}{2\pi}{\rm Re}[Z(\omega)]\tilde\Psi(\omega)  \, ,
\end{eqnarray}
where ${\rm Re}[Z(\omega)]=R/[1+(\omega/\omega_{\rm c})^2]$, and $\tilde{\Psi}(\omega)$ is defined by a digamma function~\cite{Leppakangas2008}.
The last (imaginary) contribution is for practical systems, with finite temperatures, of the same size as the real part: It stays small for environments inducing weak transition rates for the lab-frame qubit, which we assume to be true in this article.
In more details, this extra contribution is assumed to be small compared to the anharmonicity of the qubit.
The other (and possibly large) imaginary term is independent of the resistance at usual frequencies which are well below $\omega_{\rm c}$ and produces a constant $ -  \mathrm{i}/2C_{\rm int}$.
As this enters to a master equation through the matrix elements of $\hat Q_{\rm int}$,
one obtains finally a coherent renormalization term $Q^2_{\rm int}/2C_{\rm int}$, as obtained also in Eq.~(\ref{eq:TransmonRenormalization}).
This is the desired result.
In the same way, such consistency of the capacitance renormalization between the two approaches can also be shown to hold for the circuit of Fig.~\ref{fig:SimpleImpedances}(b) with counter-term as in Eq.~(\ref{eq:RNFigre1ub}).
The analysis of this section also holds exactly for $C> 0$.


\subsection{Parameters $\Delta$ and $q_0$ for a transmon qubit}\label{sec:SpinBosonConnectionLab}

After theoretically indentifying the capacitance renormalization caused by the environment to the superconducting artificial atom,
we do the reduction of the transmon to a two-level system using Hamiltonian $\hat H_{\rm tr}^0$, Eq.~(\ref{eq:RenormalizationFormal}).
We can now make a connection between the parameters of the transmon qubit and the spin-boson parameter $q_0$.

For typical transmon parameters, the energy-level difference between the ground and the first excited state is
\begin{eqnarray}
\Delta &\approx& \frac{1}{\hbar}\sqrt{8E_{\rm J}E_C} \, .
\end{eqnarray}
Here, for example, for Hamiltonian of Eq.~(\ref{eq:TransmonHamiltonianCorr2})
the charging energy $E_C=e^2/2(C_{\rm J}+C_{\rm g}^0)$.
The transmon is practically a non-linear resonator, which reduces to a two-level system when maximally only two lowest energy levels are populated.
The relevant quantity describing this reduction is the anharmonicity (difference between the first and the second energy-level differences),
\begin{eqnarray}
\hbar\Delta_{\rm an}=E_2-E_1-(E_3-E_2)\approx E_C \, .
\end{eqnarray}
This variable will play an important role in a practical realization, since the drive amplitudes $\Omega_i$ of Eq.~(\ref{eq:DrivingTerms}) need to be smaller than the
non-linearity of the qubit, 
as discussed in Sec.~\ref{sec:ErrorEstimation}

The transverse matrix element of the operator $\hat Q$ is on the other hand
\begin{eqnarray}
\vert\langle \downarrow \vert \hat Q \vert \uparrow \rangle\vert^2 &=&  e^2\sqrt{\frac{E_{\rm J}}{2E_C}} =  e^2\frac{R_{\rm Q}}{\pi Z_{\rm J}} \, .\label{eq:TwoLevelSigma}
\end{eqnarray}
Applying the result of Eq.~(\ref{eq:TwoLevelSigma}), and comparing to the form of the spin-boson Hamiltonian of Eq.~(\ref{eq:SpinBosonOur}),
we get the connection
\begin{eqnarray}\label{eq:q0Variable1}
\hat Q_{\rm int}&=& \beta\hat Q=\beta e\sqrt{\frac{R_{\rm Q}}{\pi Z_{\rm J}}}\hat\sigma_x \nonumber\\
&\equiv& \frac{q_0}{2}\hat\sigma_x \, ,
\end{eqnarray}
where now
\begin{eqnarray}\label{eq:Connectionq0}
q_0= 2e\beta \sqrt{\frac{R_{\rm Q}}{\pi Z_{\rm J}}} \, .
\end{eqnarray}
Here again $\beta=C_{\rm g}/(C_{\rm J}+C_{\rm g})$, where the ground capacitance is the unnormalized (original) one, $C_{\rm g}$,
whereas in the definition of the charging energy
and system energy levels the effective ground capacitance $C_{\rm g}^0$ appears.

\subsection{Parameter $\alpha$ for a transmon qubit (ohmic spectral density)}\label{sec:KondoParamter}

A central situation in the spin-boson theory is the case of an ohmic environment. 
Assuming an ohmic impedance,  
${\rm Re}[Z_{\rm eff}]=R$, we have $J(\omega)=R\omega \equiv \eta \omega$. This yields a Kondo parameter
\begin{eqnarray}
\alpha=\frac{1}{\pi} \beta^2 \frac{R}{Z_{\rm J}} 
\end{eqnarray}
The coupling $\alpha$ scales linearly with $R$ and is reduced by the capacitive shunting by the ground capacitance ($\beta<1$).
The relevant quantity to compare $R$ is the characteristic impedance of the Josephson junction, $Z_{\rm J}$.
The size of $\alpha$ when realized in the rotating frame is studied in Sec.~\ref{sec:ActualDesign}.

Moreover, for a transmon qubit and
for $\alpha\ll 1$ (weak-coupling limit) there is a direct connection between $\alpha$ and the quality factor of the qubit.
A golden rule calculation gives here for the decay rate~\cite{Shnirman2002} (inverse quality factor)
\begin{eqnarray}
\frac{\Gamma_{\downarrow}}{\Delta}  = \beta^2 \frac{R}{Z_{\rm J}}=\pi \alpha \, .
\end{eqnarray}
The limit $\beta=1$ (no shunting of voltage fluctuations) is the result for a dissipative classical resonator.
This direct connection appears since we have treated the transmon as a harmonic oscillator, with weak non-linearity, which is a good approximation
since $E_{\rm J}\gg E_C$.
The relation between the energy decay rate $\Gamma_{\downarrow}$ and the spin-boson parameter $\alpha$ has been studied recently in Ref.~\cite{Magazzu2017}
in the case of a high-anharmonicity flux qubit coupled to an open transmission line.

We note that if we would consider the Cooper-pair box qubit, working in the limit $E_{\rm J}\ll E_C$,
we  would have $q_0=2e$, leading to $\alpha=R/R_{\rm Q}$.
There, a resistance $R=R_{\rm Q}$  is then needed to reach $\alpha=1$.

\section{Tailoring an ohmic bath in the rotating frame}\label{sec:ActualDesign}
In this section,
we consider constructing an ohmic bath in the rotating frame from multiple microwave resonators with broadening. 
Each such resonator can be, for example, a superconducting lumped element $LC$ resonator integrated with a resistive element $R$,
 or a superconducting coplanar resonator with a leakage to an open transmission line.
After a qualitatively analysis of the achievable Kondo parameter $\alpha$, Sec.~\ref{sec:QualitativeAnalysis},
we introduce our method and show a numerical example of the bath construction, Sec.~\ref{sec:EngineeringSpectralDensity}.
Analytical relations for bath properties are derived in Sec.~\ref{sec:AnalyticalRelations} and
robustness against parasitic coupling between neighboring resonators is analyzed in Sec.~\ref{sec:Robustness}.


\subsection{Ohmic spectral density in the rotating frame}\label{sec:QualitativeAnalysis}

Let us first apply the idea presented in Sec.~\ref{sec:EffectiveHamiltonian} to realize an effective ohmic environment in the rotating frame.
We first note that in our effective system
\begin{eqnarray}
\omega_{\rm c}\ll\omega_1 \, ,
\end{eqnarray}
where $\omega_1$ is the dominant Rabi frequency, which is tuned to the energy of the superconducting qubit, $\Delta\sim 2\pi\times 7$~GHz,
and the cut-off frequency $\omega_{\rm c}\lesssim 2\pi\times 100$~MHz. This means that
we practically need a linearly increasing impedance to create a linearly increasing $J_{\rm eff}(\omega)$,
since here $J(\omega)=\omega {\rm Re}[Z(\omega)]\approx \omega_1 {\rm Re}[Z(\omega)]$.

\begin{figure}
\includegraphics[width=\columnwidth]{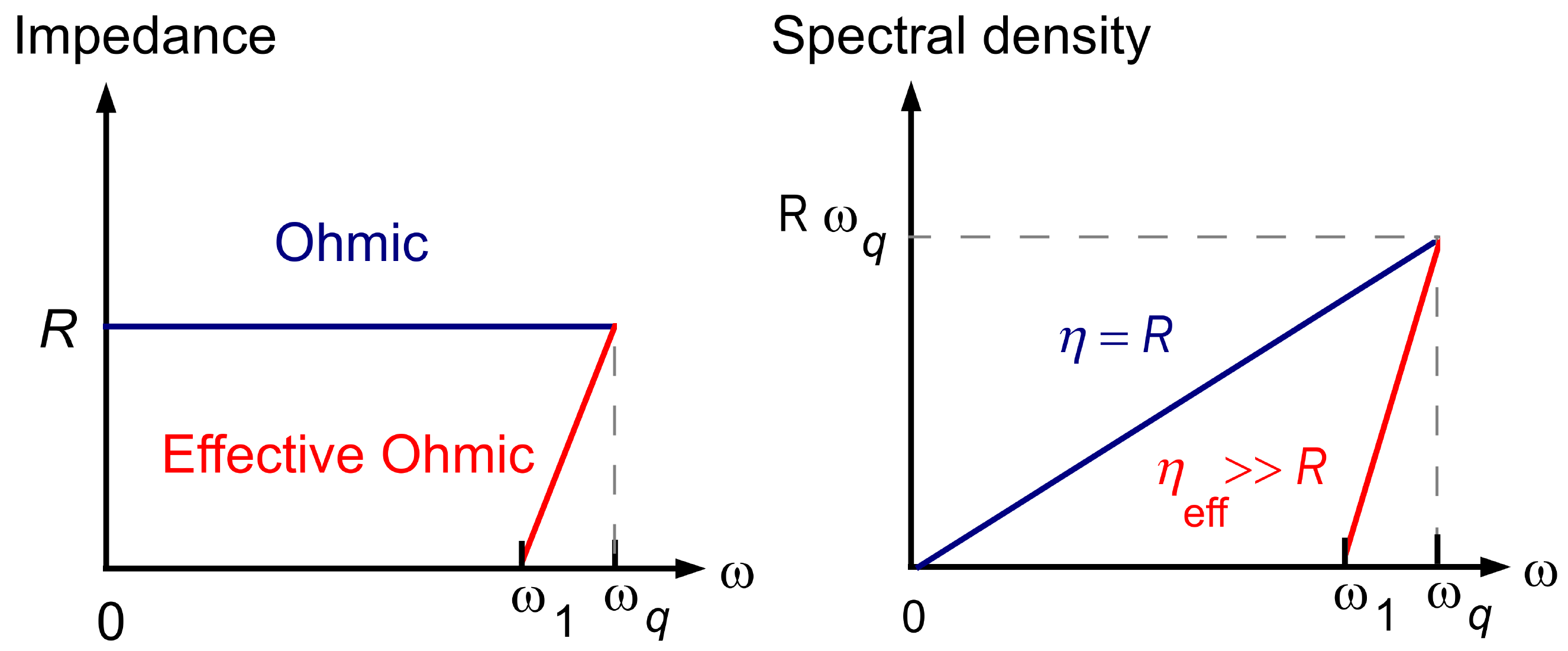}
\caption{Qualitative forms of the impedance ${\rm Re}[Z(\omega)]$ and spectral density $J(\omega)$ of two different environments,
one being ohmic in the laboratory frame (blue lines) and one being ohmic in the rotating frame (red lines).
For the same value of impedance at certain frequency $\omega_{ q}\gtrsim\omega_1$, ${\rm Re}[Z(\omega_{ q})]=R$, the coupling parameter $\eta=\partial J(\omega)/\partial\omega$ can be essentially larger in the rotating frame.
} \label{fig:SpectralDensities}
\end{figure}

Let us now assume that a parameter $R=\eta$ in some ohmic environment of the original system
describes also the maximum value of the spectral density in the constructed effective system.
Practically, such a parameter corresponds to a characteristic impedance of the microwave transmission line or resonator.
In this discussion, for simplicity, we neglect the factor 4 difference between the laboratory-frame and the rotating-frame spectral densities.
Let us denote $\omega_{ q}$ as the frequency where the maximal impedance is reached in the effective system and the two impedances meet,
so we have $J(\omega_{ q})=R\omega_{ q}$, as depicted in Fig.~\ref{fig:SpectralDensities}.
This gives for the coupling parameter in the rotating frame
\begin{eqnarray}\label{eq:EffectiveEta}
\eta_{\rm eff}=R\frac{\omega_{ q}}{\omega_{ q}-\omega_1}= R\left(1+\frac{\omega_{1}}{\omega_{\rm c}}\right) \, .
\end{eqnarray}
We see that establishing a linear increase of $J(\omega)$ in the rotating frame, we can realize
an essentially larger $\eta_{\rm eff}$, with the same maximal impedance $R$.
It can also be interpreted that the impedance of the environment is effectively increased,
without a change in the material design.

By applying this idea for a system with a transmon qubit we then get for the effective coupling in the rotating frame (accounting for the factor 4)
\begin{eqnarray}\label{eq:NoiseCorrelatorNew2}
\alpha_{\rm eff} &=&\frac{\beta^2}{4\pi}  \frac{\eta_{\rm eff}}{Z_{\rm J}} = \frac{\beta^2}{4\pi} \frac{R}{Z_{\rm J}}\left(1+\frac{\omega_{1}}{\omega_{\rm c}}\right) \, .
\end{eqnarray}
The individual multiplied contributions play an important role in determining the magnitude of $\alpha_{\rm eff}$.
The term $\beta^2/4\pi$ reduces the coupling at least by an order of magnitude.
Also the (maximal) resistivity needs to be relatively small, $R/Z_{\rm J}< 1$.
If we assume that these two contributions reduce the coupling by two-to-three orders of magnitude, then (in this example)
it is the role of the term $1+\omega_1/\omega_{\rm c}\approx \omega_1/\omega_{\rm c}$ to counteract this contribution.
For example, we would need $\omega_1/\omega_{\rm c}\approx 10^2$ in order to reach very strong couplings $\alpha_{\rm eff}\sim 0.1-1$.
This corresponds to a relatively narrow-bandwidth environment, $\omega_{\rm c}\lesssim 2\pi\times 100$~MHz. This qualitative demand should be considered together with the restriction to drive strengths $\Omega_1 $ that are much weaker than the transmon qubit anharmonicity, $\Delta_{\rm an}\lesssim 2\pi \times \SI{350}{MHz}$, and that
the Rabi frequency has to be above the cut-off of the effective environment, $\Omega_1>\omega_{\rm c}$, see Sec.~\ref{sec:ErrorEstimation}.

\begin{figure}
\includegraphics[width=\columnwidth]{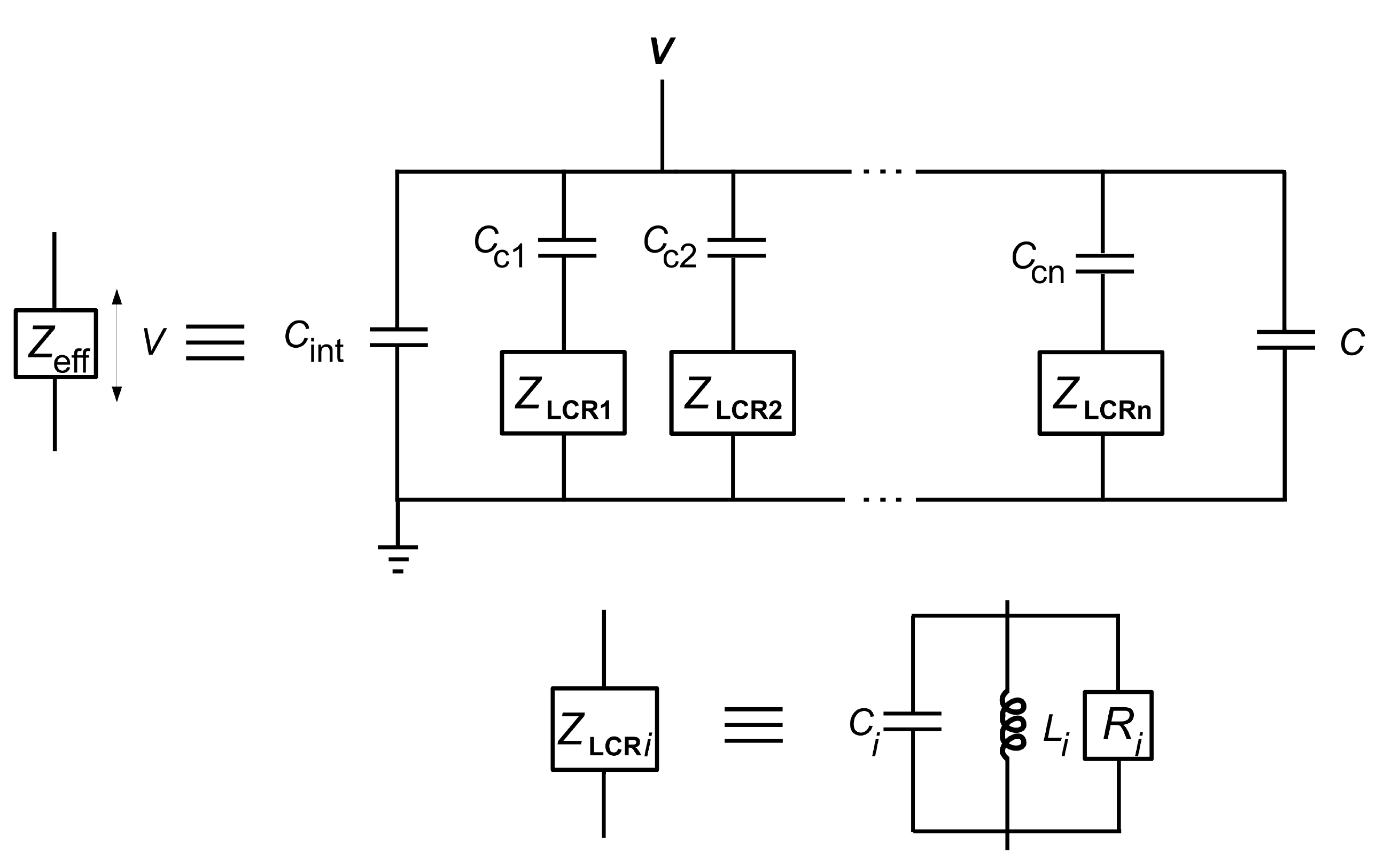}
\caption{We consider constructing the bosonic environment from multiple $LCR$ resonators coupled capacitively to a superconducting qubit.
Each resonator can be a superconducting lumped element $LC$ resonator integrated with a resistive element $R$
or, for example, a superconducting coplanar resonator with leakage to an open transmission line.
The qubit itself contributes to the effective impedance through the interaction capacitance $C_{\rm int}$, Eq.~(\ref{eq:InteractionCapacitance}).
The resonators are also assumed to be in parallel with an extra capacitor $C$, describing the coupling of the qubit antenna to ground.} \label{fig:Impedance}
\end{figure}

\subsection{Bath engineering with multiple resonators}\label{sec:EngineeringSpectralDensity}
In this work, we consider constructing the environmental impedance by using a set of $LCR$ resonators,
each of them coupled through a coupling capacitor $C_{{\rm c }i}$, as shown in Fig.~\ref{fig:Impedance}.
We desire a method that is based on a feasible manipulation of resonator parameters.
Possible methods for tailoring the spectral density are varying the individual couplings of the resonators to the qubit 
and varying the spacings between the resonance frequencies. A general recipe that can be implemented in an experiment is the following:
\begin{itemize}

\item Realize all resonators with slightly different frequencies, by varying their inductances $L_i$ and/or capacitances $C_i$. 

\item Shape the spectral function by changing individual coupling capacitances $C_{{\rm c}i}$ and/or resonance-frequency spacing.

\end{itemize}
The resonator broadenings, defined by variables $R_i$, can be used to shape the spectral function of the bath as well, but more importantly,
it is closely connected to the achievable Kondo parameter $\alpha$, as shown below.

\begin{figure}
\includegraphics[width=0.9\columnwidth]{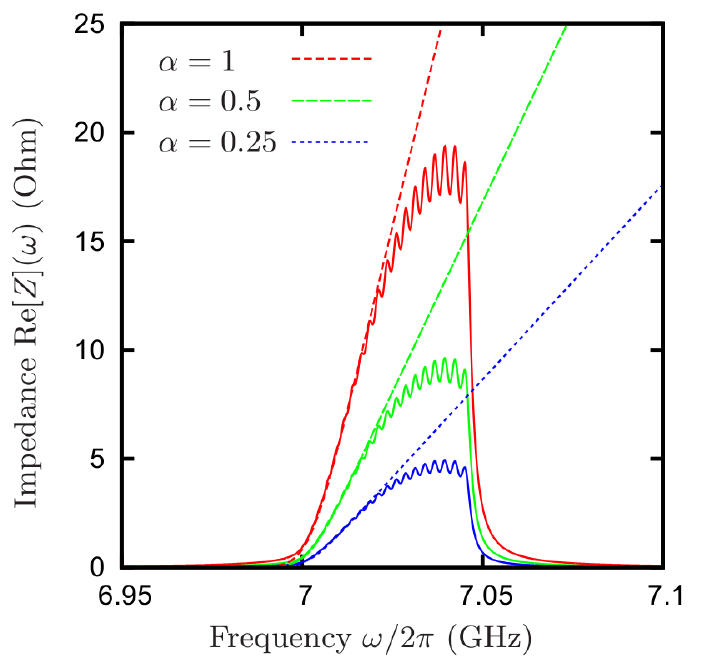}
\caption{Effective ohmic spectral density with three different Kondo parameters $\alpha$ in the rotating frame at $\omega_1/2\pi=7$~GHz.
The impedance is constructed from $N=20$ dissipative resonators with internal $Q\approx 2.2\times 10^3$.
A linear decrease in bath impedance ${\rm Re}[Z(\omega)]$ is obtained here by
reducing the coupling capacitance from $0.5$~fF quadratically to zero ($\sim 1-(i-1)^2/N^2$, where $i$ is the number of the resonator),
while increasing the inductance linearly (with $i$). Different couplings $\alpha$ correspond to different parallel capacitors $C$, such that $C+C_{\rm int}$ takes the values $70$~fF ($\alpha=1$), $\sqrt{2}\times 70$~fF ($\alpha=1/2$), and $2\times 70$~fF ($\alpha=1/4$).
The used transmon parameters are $Z_{\rm J}=200$~$\Omega$, $\beta=1/\sqrt{2}$ and resonator $Z_{LC}\approx 113$~$\Omega$}\label{fig:NumericalImpedance}
\end{figure}

A practical example of bath shaping using our approach is shown in Fig.~\ref{fig:NumericalImpedance},
where an effective ohmic impedance is constructed from $N=20$ resonators
by varying inductances $L_i$ and coupling capacitances $C_{{\rm c}i}$.
A straightforward method for calculating the total impedance (and thereby the spectral function) of similar circuits is given in Appendix~C.

\subsection{Analytical relations}\label{sec:AnalyticalRelations}
More fundamental connections between the chosen parameters
and the achievable spectral density exists.
Below, we first show analytically how the broadening and coupling of individual modes relate to $\alpha$.
After this we consider explicit formulas for the the size of the individual couplings and
study how the size of the constructed (smooth) impedance depends on resonator properties and the resonance-frequency density.
We also estimate the size of the transmon-capacitance renormalization.

\subsubsection{Kondo parameter $\alpha$}\label{sec:KondoAnalytically}

Let us first analyze how the linear increase of spectral density relates to the coupling to individual broadened resonators.
It is reasonable to assume that the steepness of the spectral density at low effective frequencies  (see for example Fig.~\ref{fig:NumericalImpedance})
is similar to, or limited by, the spectral steepness related to the individual broadened resonators.
The following discussion is made for a laboratory-frame system, but the qualitative result is independent of the chosen frame.

We then evaluate the decay rate of the qubit due to single environmental broadened resonator. According to the golden rule, the decay rate is
\begin{eqnarray}
\Gamma =\frac{\gamma}{\gamma^2+4\omega^2}g^2 \, ,
\end{eqnarray}
where $\gamma$ is the width (decay rate) of the resonator, $\omega$ the frequency with respect to the resonance frequency,
and $g\equiv q_0g_i/\hbar$ the total coupling.
The result for the decay rate is strictly valid for small couplings $g\ll\gamma$,
but this formula indeed provides a general connection between an individual resonator spectral density and a coupling to the qubit.
The derivative of the golden rule decay rate is
\begin{eqnarray}
\frac{\partial\Gamma}{\partial\omega} &=&  - 8\left(\frac{g}{\gamma}\right)^2 \frac{\frac{\omega}{\gamma}}{\left[1+4\left(\frac{\omega}{\gamma}\right)^2\right]^2} \, .
\end{eqnarray}
This has a maximal value $\gtrsim (g/\gamma)^2$.
Assuming that we synthesize a linear increase of the spectral density which qualitatively follows this steepness,
we can relate this directly to the parameter $\alpha$,
\begin{eqnarray}
\frac{\partial\Gamma}{\partial\omega}=\pi\alpha\sim \left(\frac{g}{\gamma}\right)^2 \, .
\end{eqnarray}
We then find that for couplings $\alpha\sim 1$ at least some of the resonators are in the strong-coupling regime ($g\sim \gamma$).
It is, however, not needed that individual resonators are in the ultra-strong coupling regime.
This seemingly fundamental result
states that the onset of the single-resonator strong-coupling regime, which comes together
with non-Markovian system-environment interaction, is closely related to the strong-coupling in the spin-boson model ($\alpha\sim 1$),
when the environment is constructed from multiple resonators.

\subsubsection{Coupling to individual resonators}
Let us consider now how the coupling to an individual resonator, $g$, relates to the system parameters.
A Hamiltonian for the qubit coupled to a single (non-dissipative) resonator with inductance $L_1$ and capacitance $C_1$ is here of the form
\begin{eqnarray}
H&\approx&\frac{\hbar\Delta}{2} \hat \sigma_z + \hbar\omega_1 \hat b^\dagger \hat b \\
&-& \frac{1}{2}\beta\frac{C_{\rm c}}{C+C_{\rm int}} \sqrt{\frac{C_{\rm T}}{C_1}} \hbar\sqrt{\Delta\omega_1} \left( \hat b^\dagger - \hat b \right)\left( \hat \sigma_+ - \hat \sigma_- \right) \, , \nonumber
\end{eqnarray}
where $\hbar\Delta=\sqrt{8E_{\rm J}E_C}$, $E_C=e^2/2C_{\rm T}$, $C_{\rm T}=C_{\rm J}+CC_{\rm g}/(C+C_{\rm g})$, $\omega_1=1/\sqrt{L(C_1+C_{\rm c})}$,
and we have assumed that $C_{\rm c}\ll C_{\rm J},C_1,C_{\rm g}$.
For equal system frequencies, $\Delta=\omega_1$, we get
\begin{equation}\label{eq:CouplingToSingleResonator}
g= \beta\frac{C_{\rm c}}{C+C_{\rm int}} \sqrt{\frac{C_{\rm T}}{C_1}} \Delta \, .
\end{equation}
A practical example is $\beta=1/\sqrt{2}$, $C_{\rm c}=0.1$~fF, $C+C_{\rm int}=70$~fF, $C_1=2C_{\rm T}=200$~fF, and $\Delta/2\pi=7$~GHz, which gives $g/2\pi=5$~MHz. In the rotating frame the coupling is halved to $2.5$~MHz

When constructing the spectral density using multiple resonators with internal losses, the coupling to individual resonators is reduced.
This is due to the collective capacitance due to all other resonators
\begin{eqnarray}
C_{\rm c}^{\rm total}\equiv\sum_{i=1}^N C_{{\rm c}i} \, .
\end{eqnarray}
We assume here $N\gg 1$ so that the considered resonator can be included in the sum with negligible error.
The effect to coupling to a single resonator is the same as increasing the extra capacitance to ground as $C\rightarrow C+C_{\rm c}^{\rm total}$. The coupling to qubit is then approximately
\begin{equation}\label{eq:CouplingToSingleResonator2}
g= \beta\frac{C_{\rm c}}{C+C_{\rm int}+C_{\rm c}^{\rm total}} \sqrt{\frac{C_{\rm T}}{C_1}} \Delta \, .
\end{equation}
Here also  $\hbar\Delta=\sqrt{8E_{\rm J}E_C}$, $E_C=e^2/2C_{\rm T}$, but now with $C_{\rm T}=C_{\rm J}+(C+C_{\rm c}^{\rm total})C_{\rm g}/(C+C_{\rm c}^{\rm total}+C_{\rm g})$.


\subsubsection{Value of constructed smooth impedance}\label{sec:InternalShunting}
Let us now study the size of impedance synthesized within our method.
For this we consider first establishing a rectangular impedance
between certain frequencies $\omega_1$ and $\omega_1+\omega_{\rm interval}$.
Two parameter-limits lead to simple analytical formulas: (i) when the collective coupling  $C_{\rm c}^{\rm total}\gg C_{\rm int}+C$ 
and (ii) when $C_{\rm c}^{\rm total}\ll C_{\rm int}+C$.

Let us first consider the case $C_{\rm c}^{\rm total}\gg C_{\rm int}+C$.
In this case, the effective parallel capacitive shunting is not due to the qubit or the capacitance $C$, but due to all the other resonators.
Here, assuming the same coupling capacitance $C_{{\rm c}i}=C_{{\rm c}}$ for all $N$ resonators, we get a reduction of the impedance seen by the qubit of one resonator
(due to shunting of the other resonators) by a factor
\begin{eqnarray}
\left(\frac{C_{\rm c}}{NC_{\rm c}}\right)^2=\left(\frac{1}{N}\right)^2 \, .
\end{eqnarray}
Each resonator contributes to the real part of the effective impedance (before the considered reduction) with a Lorentzian of area $\omega_i Z_{LCi}$ and a width $\delta\omega=\omega_{i} R_i/Z_{LCi}$, where the characteristic impedance of resonator $i$
is
\begin{eqnarray}\label{eq:CharacteristicImpedanceResonator}
Z_{LCi}=\sqrt{\frac{L_i}{C_i}} \, .
\end{eqnarray}
Then, the average value of the real part of the environmental impedance is (assuming a nearly constant characteristic impedances and an interval $\omega_{\rm interval}\ll \omega_1$)
\begin{eqnarray}
&&R\approx\frac{1}  {\omega_{\rm interval}}  \int_{\omega_1}^{\omega_1+\omega_{\rm interval}} {\rm Re}[Z(\omega)]d\omega \\
&\approx& \frac{1}{\omega_{\rm interval}}\left(\frac{1}{N}\right)^2\times N\omega_1 Z_{LC}=Z_{LC}\frac{\omega_1}{\omega_{\rm interval}}\frac{1}{N} \, . \nonumber
\end{eqnarray}
We see that a fundamental limit is set by the characteristic impedance of the resonators.
Indeed, for a typical set of parameters we find numerically that
\begin{eqnarray}
R\sim Z_{LC} \, .
\end{eqnarray}
However, the effect of such impedance to the transmon is actually large, since the transmon impedance is usually of the same magnitude,
which is not the regime we want to be in.

In the case $C_{\rm c}^{\rm total}\ll C_{\rm int}+C$, the effective parallel shunting is due to the qubit contribution $C_{\rm int}$ and capacitance $C$.
This is practically the regime of our proposed system.
Here, the preceding results are valid with an additional reduction factor $[N C_{{\rm c}}/ (C_{\rm int}+C)]^2$.
We then estimate for the (rectangular) impedance achieved by the considered method,
\begin{eqnarray}
R\sim \left( \frac{N C_{{\rm c}}}{ C_{\rm int}+C} \right)^2 Z_{LC}\frac{\omega_1}{\omega_{\rm interval}}\frac{1}{N} \, . 
\end{eqnarray}
In this regime,
the size of the dimensionless coupling parameter $\alpha$ can then be controlled by 
the capacitance $C$, which is done in the simulation of Fig.~\ref{fig:NumericalImpedance}.

\subsubsection{Transmon capacitance renormalization}
The environmental impedance can affect the qubit parameters through a capacitance renormalization.
Applying the approach described in Sec.~\ref{sec:GeneralCircuitHamiltonian}, we identify
the Hamiltonian of the circuit when resistances are put to zero. Here the parallel $LCR$ circuits becomes effectively shorts.
The environmental capacitance as seen by the qubit is then 
\begin{eqnarray}\label{eq:DesignCircuitRenormalizationC}
C_{\rm env}=C_{\rm c}^{\rm total} +C\,.
\end{eqnarray}
This fully coherent system corresponds to the Hamiltonian of Eq.~(\ref{eq:TransmonHamiltonianCorr2}) with 
\begin{eqnarray}\label{eq:DesignCircuitRenormalizationC2}
C^0_{\rm g}=\left[ C_{\rm g}^{-1} + \left(C_{\rm env}\right)^{-1}\right]^{-1} \, .
\end{eqnarray}
Depending on the size of the term of Eq.~(\ref{eq:DesignCircuitRenormalizationC})
in comparison to $C_{g}$, the contribution of this can be significant (for example when $C_{\rm env}<C_{\rm g}$ and $C_{\rm J}\sim C_{\rm g}$).
This correction then needs  to be included in the free Hamiltonian, Eq.~(\ref{eq:TransmonHamiltonianCorr2}).

We note that
the same result is obtained also by reducing the characteristic impedance of resonators to zero, by taking the resonator capacitances to infinity.
This gives for the effective capacitance of each resonator lead $C_{{\rm c}i}$ and thereby again
an effective environmental capacitance as in Eq.~(\ref{eq:DesignCircuitRenormalizationC}).


\subsection{Robustness against parasitic coupling}\label{sec:Robustness}

\begin{figure}
\includegraphics[width=0.9\columnwidth]{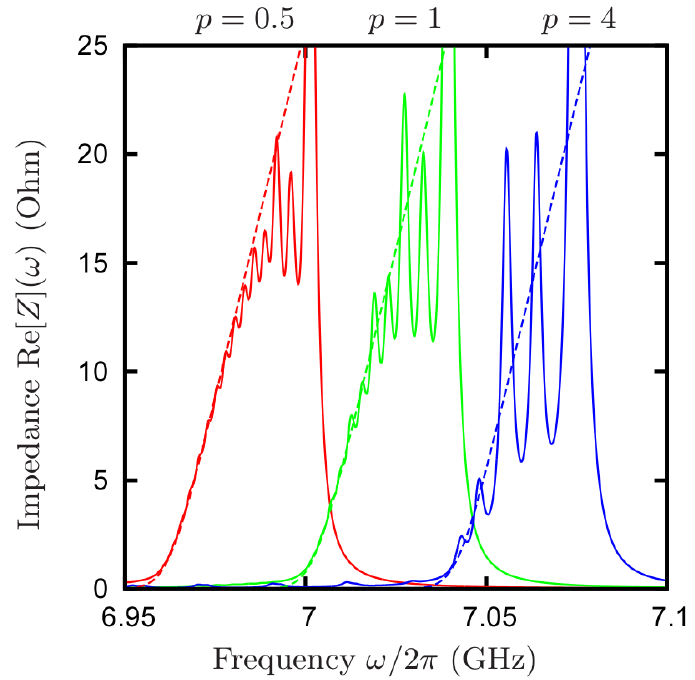}
\caption{The effect of parasitic resonator-resonator coupling to the impedance of system in Fig.~\ref{fig:NumericalImpedance}.
Here $p=C_{\rm p}/C_{i}^{\rm max}$ corresponds to the relative strength of the parasitic coupling,
where $C_{\rm p}$ is the nearest-neighbor parasitic capacitance and $C_{i}^{\rm max}=0.5$~fF is the maximal coupling between a resonator and qubit.
The other parameters are as in Fig.~\ref{fig:NumericalImpedance} for $\alpha=1$.
The curves have been separated by $0.04$~GHz and the dashed lines correspond to the spectral densities with $\alpha=1$.
We find that the low-frequency part of the impedance is practically unchanged when parasitic coupling is of the same magnitude or less than the (maximal) qubit-resonator coupling. 
}\label{fig:NumericalImpedanceParacitic}
\end{figure}

Here, we numerically study deviations in the spectral function due to a parasitic mutual coupling of bosonic bath resonators.
As described in Appendix~C, we assume a capacitive nearest-neighbor coupling between resonators, with cyclic boundary conditions.
Unwanted substructure that is introduced by this mutual coupling is suppressed when resonators nearby in frequency are arranged also spatially adjacent
(except at the boundary from the largest to the smallest).
A numerical simulation of the effect of parasitic coupling
is provided in Fig.~\ref{fig:NumericalImpedanceParacitic}.
We generally find that the resonator-resonator coupling should be of the same order or less than the coupling of individual resonators, so that
our construction method works.
In the opposite limit, the individual peaks are pushed away from each other and become visible.
We can then summarize two important findings for tailoring the impedance for Kondo couplings $\alpha\sim 1$ using the presented method:
\begin{itemize}

\item Coupling between the qubit and at least some of the resonators has to be in the strong-coupling limit, $g\sim \gamma$.

\item Parasitic coupling between resonators should be maximally of the same order as the maximal coupling to the qubit, $g$.

\end{itemize}

\section{Experimental realization}\label{sec:Experiment}

In this section, we provide a brief description of an experimental realization of a spin-boson quantum simulator based on a modular flip-chip approach. In addition, we discuss experimental protocols that allow one to access interesting quantities of the two-level system in the spin-boson simulator. They include the bath initialization, qubit-state preparation, and the qubit-state measurement.

\subsection{Flip-chip approach}

In our preliminary experimental realization of the spin-boson model, we place
the two-level system and the bosonic bath on two physically different chips. Both samples are mounted in a specifically designed sample box on top of each other in a flip-chip fashion \cite{Minev2016,Braumueller2018}. The qubit sample at the bottom is mounted on the ground level of the sample box, which allows for the required bond connections to the coaxial control lines, while the upper sample containing the bosonic bath is flipped upside down and therefore facing the qubit chip. The capacitive coupling between the qubit and bosonic bath is mediated via electric fields in the volume between the two samples.

We implement a bosonic bath formed by $N=20$ lumped-element resonators that individually couple to the qubit via coupling antennas. The resonators are equipped with resistive elements that allow us to tailor their internal dissipation such that they overlap in a restricted frequency band and form a bosonic bath of a smooth spectral function. A shaping of the bosonic bath impedance $Z(\omega)$ is achieved by adjusting the individual coupling strengths between qubit and the bosonic resonator modes, as described in Sec.~\ref{sec:ActualDesign}. The two-level system is formed by a concentric transmon qubit~\cite{ConcentricTransmon}, which allows for an approximately equal coupling in any direction in its plane due to its rotational symmetry.

In a preliminary experiment, we have demonstrated that the qubit decay rate can be dominated by the engineered bosonic bath in a spectral range of $\sim 500$~MHz. The bath-induced qubit decay rate at different frequencies corresponded here directly to the noise at different frequencies
in the spin-boson model ($\alpha\ll 1$), and thereby shows that quantum simulation using the flip-chip approach is possible.
A more detailed description of this experiment is provided in Ref.~\cite{Braumueller2018}.

\subsection{Measurement protocols}

In order to experimentally observe specific dynamics of the spin-boson model, we propose two possible pulse sequences that allow us to access the expectation values of $\hat \sigma _x$ [well population function $P(t)$]  as well as of $\hat \sigma _z$ (energy decay) with different bath initializations.
A brief description for the expected behavior of the well-population dynamics, function $P(t)$, is given in Sec.~\ref{sec:RelaxationDynamics}
and in Fig.~\ref{fig:OhmicRegimes}.

\subsubsection{Qubit initialization and measurement of $\langle\hat \sigma _z\rangle$ and $\langle\hat \sigma _x\rangle$}

\begin{figure} 
\includegraphics[width=\linewidth]{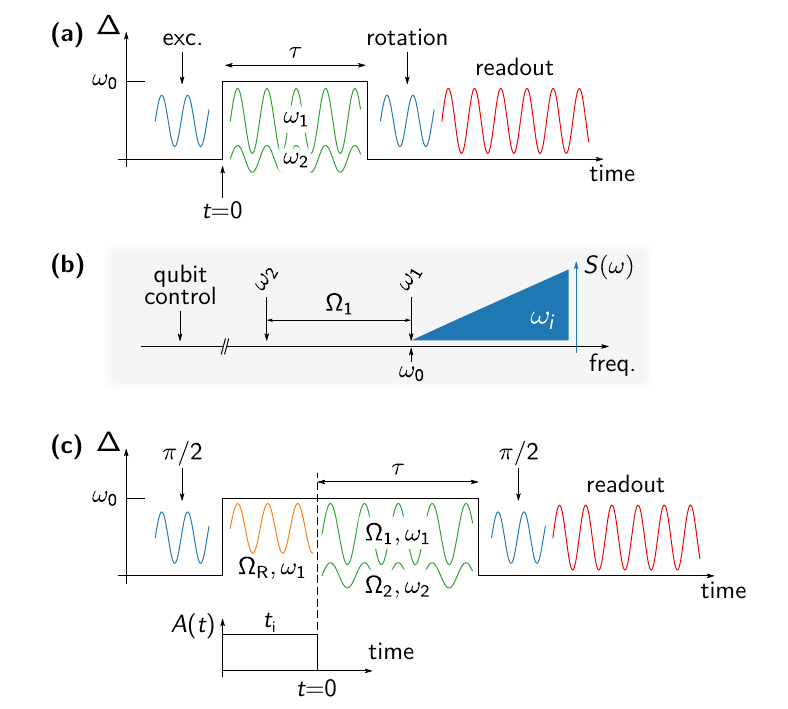}
\caption{Measurement protocols for the spin-boson simulator. (a) Pulse sequence for preparing an eigenstate of $\hat \sigma _z$ or $\hat \sigma _x$, with the qubit out of resonance with the bosonic bath, followed by interaction with the bath during time $\tau$ and readout. The qubit is tuned into the presence of the bosonic bath with a fast detuning pulse. Prior to dispersive qubit readout, we can rotate the qubit state in order to measure $\langle\hat \sigma _z\rangle$ or $\langle\hat \sigma _x\rangle$. (b) Schematic location of the drive frequencies $\omega_1$, $\omega_2$. The spectral location of the bosonic bath with individual mode frequencies $\omega_i$ is schematically depicted in blue, indicating its spectral function $S(\omega)$. (c) Proposed pulse sequence for measuring $P(t)$ including a bath initialization scheme. The qubit is initially prepared in an eigenstate of $\hat \sigma _x$ via a $\pi/2$ rotation. At $t_{\mathrm{i}}<t<0$, we initialize the bosonic bath via a strong bath drive of amplitude $\Omega_{\mathrm{R}}$ and frequency $\omega_1$. For $t>0$, we set $\Omega_{\mathrm{R}}=0$.}
\label{fig:pulsesequence}
\end{figure}
Observing the time evolution of the expectation values $\langle\hat \sigma _z\rangle$ or $\langle\hat \sigma _x\rangle$ can be performed with an extension of the measurement protocol applied in Ref.~\cite{Jochen2017}, see Fig.~\ref{fig:pulsesequence}(a-b). The qubit in the laboratory frame is initially biased to a frequency outside the spectral location of the bosonic bath. In the frequency space shown in Fig.~\ref{fig:pulsesequence}(b), this is denoted as 'qubit control'. The qubit is excited to the equatorial plane of the Bloch sphere by applying a $\pi/2$ rotation, see Fig.~\ref{fig:pulsesequence}(a). By controlling the relative phase of the successive Rabi drives~\cite{Jochen2017}, we can prepare the qubit in an eigenstate of $\hat\sigma _x$. This allows us to also initialize the effective qubit state, because at $t=0$ eigenstates remain unchanged during the transformation into the rotating frame. Alternatively, the qubit can stay in its ground state or be prepared in its excited state by applying a $\pi$ rotation prior to the start of the simulation sequence at $t=0$. With a fast frequency tuning pulse, the qubit is brought into resonance with the bosonic bath during the simulation time $\tau$, where we apply the drive tones with frequencies $\omega_1$, $\omega_2$ (see Sec.~\ref{sec:EffectiveHamiltonian}). As can be seen in the depicted pulse sequence in Fig.~\ref{fig:pulsesequence}(b), the laboratory frame qubit frequency is tuned to the lower cut-off frequency $\omega_0$ of the bosonic bath. This also corresponds to zero frequency in the effective frame, given by the rotating frame frequency $\omega_1=\omega_0$. After the simulation of time $\tau$, we apply 
an optional $\pi/2$ rotation prior to qubit readout. This allows us to measure $\langle\hat \sigma _x\rangle$ of the qubit state. If no rotation is applied, we measure the qubit state along its quantization axis, $\langle\hat \sigma _z\rangle$.

\subsubsection{Bath initialization}

Within the above formalism we are able to probe the relaxation of qubit excitations for both $\langle\hat \sigma _z\rangle$ and $\langle\hat \sigma _x\rangle$. For a direct comparison with the spin-boson theory, for example presented in Ref.~\cite{Spin_Boson_Rev}, the environment has to be properly initialized in addition. On the other hand, a comparison between the results obtained using different initialization methods allows for experimentally exploring the effect of bath initialization in the spin-boson model.

In order to observe the well population function $P(t)$ as discussed in Ref.~\cite{Spin_Boson_Rev}, the qubit in the spin-boson system is initially required to be in an eigenstate of $\hat\sigma_x$ for $t<0$, with the bath being relaxed in thermal equilibrium within this condition. This can be achieved experimentally by applying a Rabi drive at the rotating frame frequency $\omega_1$ of enhanced amplitude $\Omega_{\mathrm{R}}=\Omega_1+A(t)$. After the transformation in the interaction picture, this leaves an additional term in the effective spin-boson Hamiltonian
\begin{equation}
\hat H _{\mathrm{eff}} +A(t)\hat\sigma _x,
\end{equation}
with $\hat H _{\mathrm{eff}}$ given in Eq.~\eqref{eq:EffectiveHamiltonian2}. Initialization is applied at an effective amplitude $A(t)=\Omega_{\mathrm{R}}-\Omega_1\gg g$, where $g$ is the typical coupling strength between qubit and individual bosonic mode. Figure~\ref{fig:pulsesequence}(c) shows a schematic of the proposed pulse sequence. Bath initialization takes place during $t_{\mathrm{i}}<t<0$ with $t_{\mathrm{i}}$ defined by the inverse spectral width of the bosonic bath. The simulation starts at $t=0$, where the initialization drive is switched off, $A(t)=0$, and the Rabi drives of the simulation scheme are switched on. To recover the well population function $P(t)$, we $\pi/2$ rotate the qubit state before readout in order to measure $\langle\hat \sigma _x\rangle$.

\subsection{Bath heating}


Dissipation of the bosonic bath can be implemented by adding an ancillary transmission line, providing a loss channel for bath excitations~\cite{Ioan2017}. In our approach, dissipation takes place by ohmic dissipation on-chip and therefore involves Joule heating. 
The effect gives rise to a small modification of the bath spectral function.
The main source of on-chip dissipation can be assumed to be the Rabi drive with amplitude $\Omega_1$ and frequency $\omega_1$,
which in a realistic experiment also couples directly to the bath.

If we assume that the coupling between the drive and the bath is mediated by the qubit,
the effective drive of the environment is of an approximate amplitude $(C_{{\rm c}i}/C_{\rm int})\Omega_1$ per resonator $i$.
Each resonator will couple to the drive with separate coupling.
In the system considered in Sec.~\ref{sec:ActualDesign}, each resonator has an approximative width $\gamma_i\lesssim 2\pi\times 5$~MHz.
Considering explicitly the highest-energy resonator, with the off-resonance drive $\omega_i-\omega_1\sim 2\pi\times 50$~MHz, we get an average photon number in this resonator
\begin{eqnarray}
\langle \hat n_i\rangle \approx \left(\frac{C_{{\rm c}i}}{C_{\rm int}}\right)^2\left( \frac{\Omega_1}{\omega_i-\omega_1} \right)^2\lesssim 10^{-2} \, .
\end{eqnarray}
This leads to photon dissipation rate $\Gamma_{\rm dis}=\gamma_i \langle \hat n_i\rangle \lesssim 1$~MHz.
Due to the specific form of the designed impedance, the result is approximately the same for all individual resonators.
The length of the one measurement process is roughly $1$~$\mu$s, which implies that during one measurement each on-chip resistor 
absorbs on average less than $1$ photon. The effect of this to the temperature of each resistor is small,
but can set a minimal (cooling) time interval between two successive measurement protocols.
Specific pulsing schemes that relax the environmental resonators to their ground states just before the quantum simulation can also be used~\cite{Bultink2016,McClure2016,Boutin2017}.

\section{Conclusions and discussion}\label{sec:Conclusions}

In conclusion, we have shown that a quantum simulation of the spin-boson model can be performed in a wide parameter range
using a superconducting qubit connected to a microwave circuit.
In order to probe numerically difficult parameter regimes,
we considered an extension of the driving scheme proposed in Ref.~\cite{Ballester2012}.
This effectively down-converts the system dynamics from the gigahertz to the megahertz regime,
while preserving the order of the coupling strength between the two-level system and the environment.
The approach allows for the observation of a quantum phase transition in a regime of a large effective Kondo parameter $\alpha\sim 1$,
also without the use of a high-anharmonicity superconducting qubit.
We find that this requires strong coupling between the qubit and microwave resonators in the laboratory frame.
The phase transition region in the spin-boson model corresponds to a regime with an energy decay rate of the two-level system that is comparable to its effective transition frequency.

We discussed how to experimentally probe the well population dynamics $P(t)$
under different initialization conditions of the bosonic bath. 
For this purpose, we provided concrete measurement pulse sequences, based on well-established control and detection schemes from circuit QED.
In the considered system, probing the well population dynamics corresponds to measuring the expectation value of the $\hat\sigma_x(t)$ operator. 
It is also straightforward to study other two-level system correlation functions, such as of the $\hat\sigma_z(t)$ operator, as well as the effect of bath initialization.

The proposed approach allows for engineering a rather arbitrary spectral function in a restricted frequency range.
We estimated that for a realization with a transmon qubit the spectral width of the environment must be in the range of $100$~MHz.
By controlling the drive and qubit frequencies, we can adjust the zero-frequency condition of the tailored bosonic bath, which allows us to choose the effective system temperature $T_{\rm eff}$.
By controlling the amplitude of the weaker Rabi drive, $\Omega_2$, we can tune the effective two-level system energy relative to the temperature and the cut-off frequency $\omega_{\rm c}$, which is of central importance in the spin-boson theory.
In particular, Kondo physics can be observed for an effective temperature below the Kondo temperature $T_{\rm K}$. At the Toulouse point ($\alpha=1/2$) one can estimate~\cite{KondoMicrowaves2} $k_{\rm B}T_{\rm K}\sim \hbar\Omega_2^2/\omega_{\rm c}$, which can be adjusted by $\Omega_2$.
Hence, our system can access a large parameter space of the spin-boson model via experimental drive control.
The proposed experimental approach, based on the flip-chip technique, also features a modularity that allows to probe various fabricated bosonic environments with the same qubit in successive experiments. 

\section*{Acknowledgments}
This work was supported by the European Research Council (ERC) within consolidator Grant No.
648011 and Helmholtz IVF grant 'Scalable solid state quantum computing'.
This work was also partially supported by the Ministry of Education and Science of Russian Federation in the framework of Increase Competitiveness Program of the NUST MISIS (contracts no. K2-2014-025, K2-2016-051, and K2-2016-063).
J.B. acknowledges financial support by the Landesgraduiertenf\"orderung (LGF) of the federal state Baden-W\"urttemberg and by the Helmholtz International Research School for Teratronics (HIRST).


\appendix

\section{Deriving spin-boson model parameters using an open-circuit method}

In this Appendix, we introduce the open-circuit method which can be used to model 
the considered dissipative circuits by treating the on-chip resonators by equivalent open transmission lines.
For simplicity,
we consider here the model of a Josephson junction coupled to a single dissipative $LCR$-circuit,
as shown in Fig.~\ref{fig:OpenModel}. The generalization to many resonators is straightforward but technically more involved than if determining the effective impedance classically, as done in the main part of the article.
Working with such explicit circuits helps one to check the validity of results based on more phenomenological approaches.


\subsection{Lagrangian}
\begin{figure}
\includegraphics[width=\columnwidth]{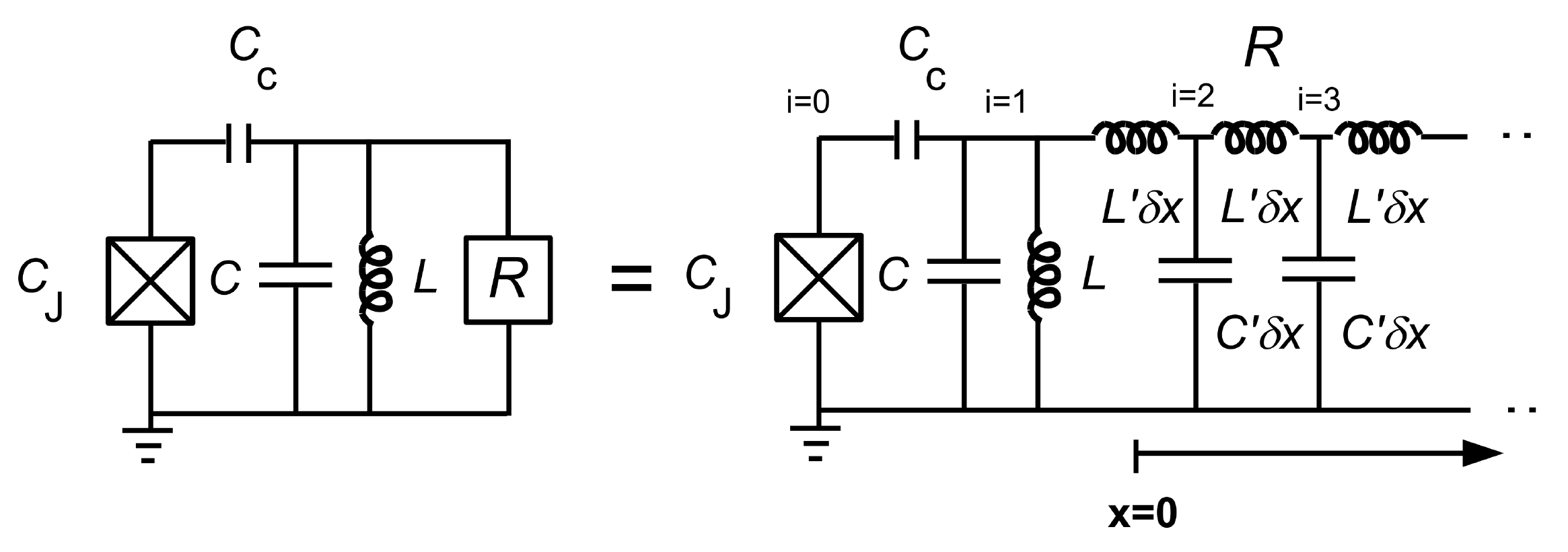}
\caption{Open-circuit model of a Josephson junction capacitively coupled to dissipative $LC$-resonator.} \label{fig:OpenModel}
\end{figure}

We consider a Josephson junction coupled to one dissipative resonator, as shown in Fig.~\ref{fig:OpenModel}.
By representing the resistor as a semi-infinite transmission line, the total Lagrangian can be written as
\begin{equation}
{\cal L}={\cal L}_{\rm env}+{\cal L}_{\rm int} + {\cal L}_{\rm JJ} \, ,
\end{equation}
where the environmental part corresponds to the Lagrangian of a semi-infinite transmission line
\begin{eqnarray}
{\cal L}_{\rm env}&=&\frac{ C\dot\Phi_1^2}{2}-\frac{\Phi_1^2}{2L}+\sum_{i \geq 2}^\infty\frac{\delta x C' \dot\Phi_i^2}{2}-\sum_{i\geq 2}^\infty\frac{(\Phi_i-\Phi_{i-1})^2}{2L'\delta x} \, .
\end{eqnarray}
The variable $\Phi_i(t)$ corresponds to the magnetic flux at node $i$ and $\dot\Phi_i$ is the corresponding voltage. The interaction part reads as
\begin{eqnarray}
{\cal L}_{\rm int}&=&\frac{ C_{\rm c}\left( \dot\Phi_{1}-\dot\Phi_0 \right)^2}{2} \, ,
\end{eqnarray}
and the Josephson-junction part
\begin{equation}
{\cal L_{\rm JJ}}=E_{\rm J}\cos\left( \frac{\Phi_{0}}{\hbar/2e} \right)+\frac{ C_{\rm J}\dot\Phi_0^2}{2}.
\end{equation}

\subsection{Hamiltonian}
Derivation of the Hamiltonian starts from the identification of the conjugated variables of the  fluxes.
These are defined as $Q_i=\partial{\cal L}/\partial \dot\Phi_i$. We get
\begin{eqnarray}
Q_0&=& C_{\rm c}(\dot\Phi_0-\dot\Phi_1)+C_{\rm J}\dot\Phi_0  \\
Q_1&=& C_{\rm c}(\dot\Phi_1-\dot\Phi_0)+C\dot\Phi_1 \\
Q_{i\geq 2}&=&\delta x C'\dot \Phi_i \, .
\end{eqnarray}
The inverse transformation has the form
\begin{eqnarray}
\dot\Phi_{i\geq 2}&=&\frac{P_i}{\delta x C'}\\
\dot\Phi_1&=&Q_1\frac{C_c+C_J}{C(C_c+C_J)+C_cC_J}+Q_0\frac{C_c}{C(C_c+C_J)+C_cC_J} \nonumber \\
&\equiv& Q_1\frac{1+C_J/C_c}{\tilde C}+\frac{Q_0}{\tilde C} \\
\dot\Phi_0&=&Q_0\frac{C_c+C}{C_J(C_c+C)+C_cC}+Q_1\frac{C_c}{C_J(C_c+C)+C_cC} \nonumber\\
&\equiv& Q_0\frac{1+C/C_c}{\tilde C}+\frac{Q_1}{\tilde C} \, .
\end{eqnarray}
Our Lagrangian corresponds to the Hamiltonian
\begin{eqnarray}
H &=& H_{\rm env}+H_{\rm int}+H_{\rm JJ}  \, ,
\end{eqnarray}
where
\begin{eqnarray}
H_{\rm env}&=& \frac{ \Phi_1^2}{2L} + \frac{Q_1^2}{2(C+C_{p1})} +\sum_{i\geq 2} \frac{Q_i^2}{2\delta xC'}\nonumber \\
&+&\sum_{i\geq 2}\frac{(\Phi_i-\Phi_{i-1})^2}{2L'\delta x}      \, .
\end{eqnarray}
Here we have defined the series capacitance as seen by the resonator, $1/C_{p1}\equiv 1/C_c+1/C_{\rm J}$.
If we identify $C_{\rm c}=C_{\rm g}$ (as in the main part of the article) we have $C_{p1}=C_{\rm int}$.
As the Lagrangian and Hamiltonian terms for $Q_0$ are the same as for $Q_1$, within the swap $C_{\rm J}\leftrightarrow  C_1$, we must have
\begin{eqnarray}
H_{\rm JJ}=-E_{\rm J}\cos \left(\frac{2e}{\hbar}\Phi_0\right) + \frac{Q_0^2}{2(C_J+C_{p0})}     \, ,
\end{eqnarray}
where analogously $1/C_{p0}\equiv 1/C_c+1/C$. The junction and resonator capacitances are now renormalized as expected.
The interaction term gets the form
\begin{eqnarray}
H_{\rm int}=\frac{Q_0Q_1}{\tilde C}     \,\,\,\,\,\,\,\,\,\,    ,    \,\,\,\,\,\,\,\,\,\, \tilde C=C_J+C+\frac{C_JC}{C_c} \, .
\end{eqnarray}
We note that in the main part of the article the operator $\hat Q_0$ is marked simply $\hat Q$.

\subsection{Solution}
We have now determined the form of the Hamiltonian corresponding to the circuit of Fig.~\ref{fig:OpenModel}.
The next step is to establish the solution when the interaction term is turned off. 
In the transmission line one obtains a wave equation whose solution can be written in the form
\begin{eqnarray}
&&\hat\Phi(x>0,t)=\sqrt{\frac{\hbar R}{4\pi}}\int_0^\infty\frac{d\omega}{\sqrt{\omega}} \\
&&\times \left[ \hat b_{\rm  in}(\omega)e^{\mathrm{i}(-k_\omega x-\omega t)}+\hat b_{\rm out}(\omega)e^{\mathrm{i}(k_\omega x-\omega t)}+{\rm H.c.} \right]\, , \nonumber
\end{eqnarray}
Here the characteristic impedance $R=\sqrt{L'/C'}$ and the wave number $k_\omega=\omega\sqrt{L'C'}$.
The bosonic terms $\hat b_{\rm in}^{\dagger}(\omega)$ and $\hat b_{\rm in}(\omega)$ correspond to incoming photon-field
creation and annihilation operators. They (as well as the out-field operators) satisfy
$\left[ \hat b_{\rm in}(\omega),\hat b_{\rm in}^{\dagger}(\omega') \right]=\delta(\omega-\omega')$ \, .

The Heisenberg equations of motion at $i=1$ read
\begin{eqnarray}
\hat{\dot{ \Phi}}_1(t) &=&  \frac{\hat Q_1}{C+C_{p1}}+ {\bf 0\times \frac{\hat Q_{0}}{\tilde C} } \label{eq:ResonatorBoundary1} \\
\hat{\dot{ Q}}_1(t) &=& -\frac{\hat \Phi_1}{L} + \frac{ \hat \Phi_{2} -\hat \Phi_1 }{\delta x L'}\nonumber\\
&\rightarrow& -\frac{\hat \Phi_1}{L} +\frac{1}{L'}\frac{\partial \hat \Phi(x=0,t)}{\partial x}. \label{eq:ResonatorBoundary2}
\end{eqnarray}
In the first equation, we have set the interaction term (boldface) to zero. The junction is now decoupled from the dissipative resonator.
These equations lead to relation
\begin{eqnarray}
&&(C+C_{p1})\hat{\ddot{ \Phi}}(x=0,t) = \nonumber \\
&&-\frac{\hat \Phi(x=0,t)}{L} +\frac{1}{L'}\frac{\partial \hat \Phi(x=0,t)}{\partial x}.
\end{eqnarray}
This is a boundary condition between the incoming and outgoing fields. The solution is obtained by Fourier transforming, which gives
\begin{eqnarray}
&&\omega^2(C+C_{p1})\left[ \hat b_{\rm in}(\omega) +\hat b_{\rm out}(\omega) \right]\\
&&=\frac{1}{L}\left[ \hat b_{\rm in}(\omega) +\hat b_{\rm out}(\omega) \right]+\mathrm{i}\frac{\omega}{R}\left[ \hat b_{\rm in}(\omega) -\hat b_{\rm out}(\omega) \right] \, . \nonumber
\end{eqnarray}
The solution is
\begin{eqnarray}
\hat b_{\rm out}(\omega)=  -\frac{1+\mathrm{i}\frac{L\omega/R}{1-(\omega/\omega_1)^2} }{1-\mathrm{i}\frac{L\omega/R}{1-(\omega/\omega_1)^2}}   \hat b_{\rm in}(\omega) \, .
\end{eqnarray}
We find that at zero as well as at infinite frequency, the boundary condition gives $\hat a_{\rm out}+\hat a_{\rm in}=0$.
Similarly, we find
\begin{eqnarray}
\hat b_{\rm in}(\omega)+\hat b_{\rm out}(\omega)&=& -2\mathrm{i}\frac{L\omega/R}{1-(\omega/\omega_1)^2-\mathrm{i}L\omega/R} \hat b_{\rm in}(\omega) \nonumber\\
& \equiv& A(\omega)\hat b_{\rm in}(\omega) \, .
\end{eqnarray}
This is proportional to the impedance of parallel $LCR$-circuit:
\begin{eqnarray}
Z_{\rm eff}(\omega) &=& \frac{1}{\frac{1}{R}+\frac{1}{\mathrm{i}\omega L}+\mathrm{i}\omega(C+C_{p1})}=\frac{R}{2}A^*(\omega) \label{eq:AppendixVoltageSolution1} \\
{\rm Re}[Z_{\rm eff}(\omega)] &=&  \frac{R}{4}\vert A(\omega)\vert^2 \, . \label{eq:AppendixVoltageSolution2}
\end{eqnarray}
The solution for the interaction voltage has the form
\begin{eqnarray}
&&V_{\rm int}(t)\equiv \frac{\hat Q_1(t)}{\tilde C}=-\mathrm{i}\frac{C_c}{C_c+C_{\rm J}}\sqrt{\frac{\hbar R}{4\pi}}\nonumber\\
&&\times\int_0^\infty d\omega \sqrt{\omega}A(\omega)\hat b_{\rm in}(\omega)e^{-\mathrm{i}\omega t} + {\rm H.c.} \label{eq:AppendixVoltageSolution3} \\
&&\left\langle V_{\rm int}(t) V_{\rm int}(0) \right\rangle_{T=0}= \left( \frac{C_c}{C_c+C_{\rm J}} \right)^2 \frac{\hbar}{\pi} \nonumber\\
&&\times \int_0^\infty d\omega \omega {\rm Re}[Z_{\rm eff}(\omega)] e^{-\mathrm{i}\omega t} \, . \label{eq:AppendixVoltageSolution4}
\end{eqnarray}
This agrees with the results given in the main part of the article.

\section{Single-mode versus continuous-mode treatment of a microwave resonator}
In this appendix, we study the connection between single- and continuous-mode treatments of a microwave resonator.
In the preceding appendix, we already derived an example of the connection between single- and multi-mode treatments,
by deriving an exact form of the amplitude function $g_i\propto A(\omega)$ in the case of a transmon coupled to a single dissipative $LCR$ resonator,
Eqs.~(\ref{eq:AppendixVoltageSolution1}-\ref{eq:AppendixVoltageSolution4}).
In this appendix, we consider relations between the single- and continuous-mode treatments by using the representation in (numerable) bosonic operators $\hat b_i$.

For simplicity, we consider here the case $q_0=2e$ (Cooper-pair box), generalization to other cases is straightforward.
In the single-mode analysis, the relative coupling strength between the qubit and a mode of frequency $\omega_{\rm r}$ (divided by the mode frequency) is given (in the absence of coupling capacitor) by
\begin{eqnarray}\label{Eq:AppendixCouplingG}
g=\sqrt{\frac{\pi Z_{LC}}{R_{\rm Q}}} \, ,
\end{eqnarray}
where $Z_{LC}$ is the characteristic impedance of a microwave resonator.
On the other hand, for the continuous mode description of the same broadened mode, it is the area of the peak that matters,
\begin{eqnarray}
\frac{q_0^2}{2\pi\hbar}\int d\omega J(\omega)=\frac{1}{R_{\rm Q}} \sum_i g_i^2=\omega_{\rm r}^2 \frac{g^2}{2}   \, ,
\end{eqnarray}
where we have used the information that in the considered case the integration over $J(\omega)$ is proportional to the characteristic impedance $Z_{LC}$.
Here the values $g_i^2$ form a peak around the central frequency $\omega_{\rm r}$, describing a broadened resonator.
We then obtain a connection between the single-mode and the continuous-mode treatments of the same peak in the spectral density
\begin{eqnarray}
g^2=\frac{2}{R_{\rm Q}}\sum_i \left(\frac{g_i}{\omega_{\rm r}}\right)^2=\frac{1}{\pi\hbar}\sum_i \left(\frac{q_0g_i}{\omega_{\rm r}}\right)^2 \, .
\end{eqnarray}
We see that the (squared) total effective strength is proportional to the sum of the squared strengths of individual modes.
Note also that in comparison to couplings $g_i$, the coupling $g$ is normalized by $\omega_{\rm r}$,
which means that it depends only on the characteristic impedance of the resonator, see Eq.~(\ref{Eq:AppendixCouplingG}).

We could also interpret such a single-mode peak as a single slice of an ohmic spectrum, at frequency $\omega_{\rm r}$, with width $d\omega$, and total coupling $g$.
This interpretation gives a relation
\begin{eqnarray}
\frac{q_0^2}{2\pi\hbar}\int d\omega J(\omega)=\omega_{\rm r}^2 \frac{g^2}{2}=\alpha\omega_{\rm r}d\omega \   \, .
\end{eqnarray}
This leads to the identification
\begin{eqnarray}
\alpha=\frac{1}{2}\frac{\omega_{\rm r}}{d\omega}g^2    \, .
\end{eqnarray}
The variable $d\omega$ is so far arbitrary, and stands here for the width of the chosen slice of the ohmic spectrum.
Also the variable $g$ is not fixed.

If we decide to fix the frequency-normalized coupling $g$ (not $\omega_{\rm r} g$), i.e.,
keep the characteristic impedances independent of frequency of the chosen slice,
then the frequency interval between resonators, $d\omega$, has to decrease with the position $\omega_{\rm r}$. This can be interpreted as
that the quality factors of individual resonators need to be identical: The resonators are equivalent up to a frequency conversion.
On the other hand, if we decide to fix $d\omega$, we obtain that the coupling needs to behave as $g^2\propto \alpha d\omega/\omega_{\rm r}$.
This increases when decreasing $\omega_{\rm r}$.
However, equivalently, the un-normalized couplings should behave as $(\omega_{\rm r}g)^2\sim g_i^2 \sim \alpha \omega_{\rm r} d\omega$.
This then shows that the actual (squared) couplings $g_i^2$ need to increase linearly with frequency, as expected.


It should be noted that
when contructing an effective bath at high frequencies instead, the contributions of individual resonator frequencies $\omega_i$ in the effective couplings $\omega_ig$ can be treated as a constant.
The bath can then be constructed by varying the resonator density, or by manipulating couplings $g$ by additional coupling capacitors,
as described in the main part of the article.

\begin{figure}
\includegraphics[width=\columnwidth]{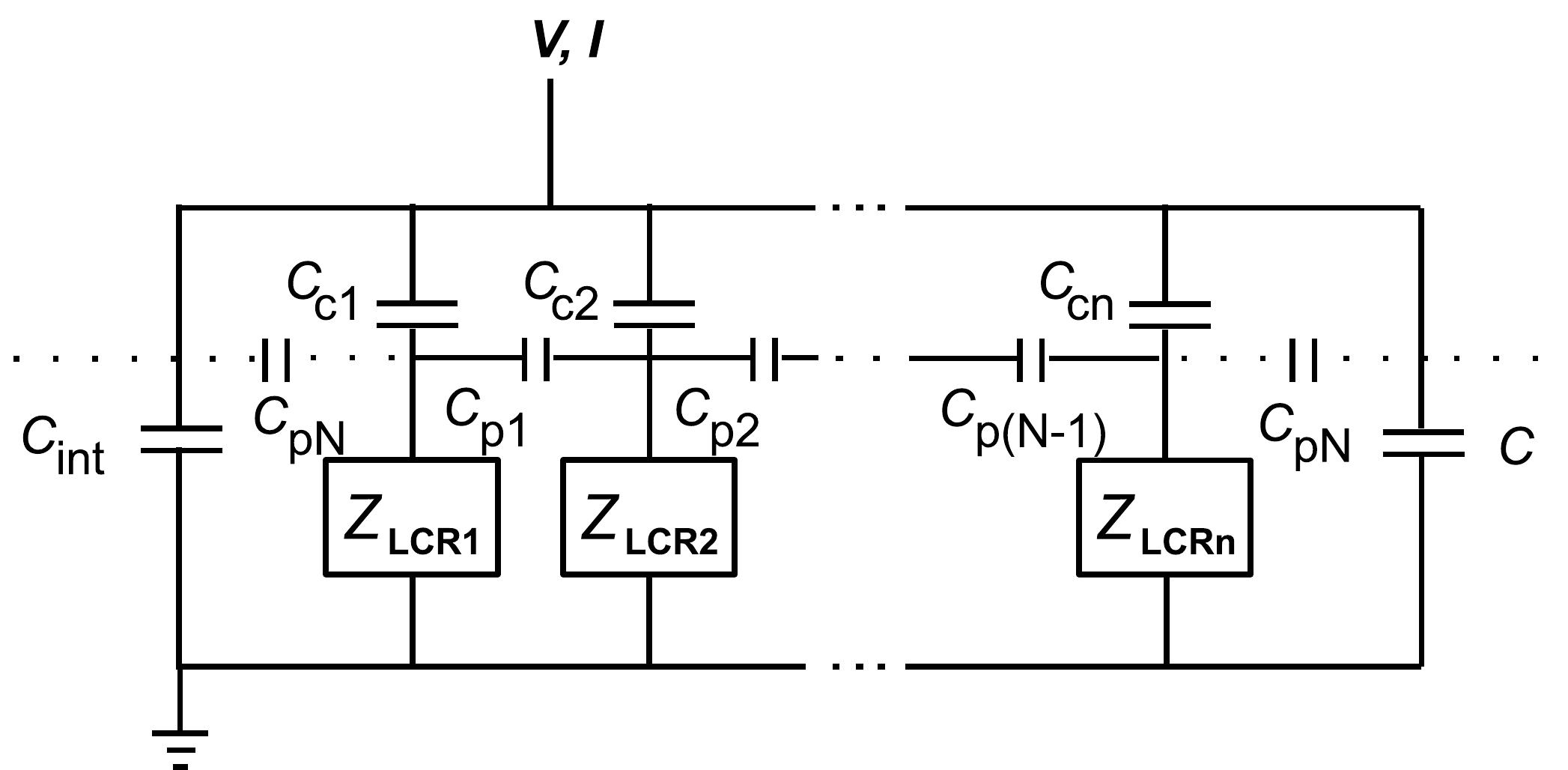}
\caption{Lumped-element model of the resonator bath with additional parasitic couplings $C_{{\rm p}i}$.} \label{fig:ImpedanceFull}
\end{figure}

\section{Determining the impedance of the environment}
We consider the generalized circuit shown in Fig.~\ref{fig:ImpedanceFull}.
We mark the voltage of island $i$, which locates between the capacitor $C_{{\rm c}i}$ and $LCRi$ element, by $V_{\rm i}$.
The impedance can then be evaluated from the conditions for the current conservation
\begin{eqnarray}
\frac{V-V_1}{Z_{\rm C1}}&=&\frac{V_1}{Z_{\rm LCR1}}+\frac{V_1-V_2}{Z_{\rm P1}}+\frac{V_1-V_N}{Z_{\rm PN}} \nonumber \\
\frac{V-V_2}{Z_{\rm C2}}&=&\frac{V_2}{Z_{\rm LCR2}}+\frac{V_2-V_3}{Z_{\rm P2}}+\frac{V_2-V_1}{Z_{\rm P1}} \nonumber \\
\ldots & &\nonumber \\
\ldots & &\nonumber \\
\ldots & &\nonumber \\
\frac{V-V_N}{Z_{\rm CN}}&=&\frac{V_N}{Z_{\rm LCRN}}+\frac{V_N-V_1}{Z_{\rm PN}}+\frac{V_N-V_{N-1}}{Z_{\rm P(N-1)}}  \, .
\end{eqnarray}
Here we represent each circuit element by their equivalent impedance, for the coupling capacitor $i$ this being $Z_{{\rm C}i}=(\mathrm{i}\omega C_{{\rm c}i})^{-1}$,
for the parasitic coupling $Z_{{\rm p}i}=(\mathrm{i}\omega C_{{\rm p}i})^{-1}$, and for the $LCR$ element $Z_{LCRi}=(\mathrm{i}\omega C_i +1/(\mathrm{i}\omega L_i)+ 1/R_i )^{-1} $.
The above set of equations can be represented as the matrix equation for island voltages $V_i$,
\begin{widetext}
\begin{equation}
\left(
\begin{matrix}
V/Z_{\rm C1} \\
V/Z_{\rm C2} \\
\ldots \\
V/Z_{\rm CN} \\
\end{matrix}
\right)
=\left(
\begin{matrix}
\frac{1}{Z_{\rm C1}}+\frac{1}{Z_{\rm LCR1}}+\frac{1}{Z_{\rm P1}} +\frac{1}{Z_{\rm PN}}  & -\frac{1}{Z_{\rm P1}} & \ldots & -\frac{1}{Z_{\rm PN}} \\
-\frac{1}{Z_{\rm P1}} & \frac{1}{Z_{\rm C2}}+\frac{1}{Z_{\rm LCR2}}+\frac{1}{Z_{\rm P2}} +\frac{1}{Z_{\rm P1}}  & -\frac{1}{Z_{\rm P2}} & \ldots \\
\ldots & \ldots & \ldots & \ldots \\
-\frac{1}{Z_{\rm PN}} & \ldots & -\frac{1}{Z_{\rm P(N-1)}} & \frac{1}{Z_{\rm CN}}+\frac{1}{Z_{\rm LCRN}}+\frac{1}{Z_{\rm PN}} +\frac{1}{Z_{\rm P(N-1)}}
\end{matrix}
\right)
\left(
\begin{matrix}
V_1 \\
V_2 \\
\ldots \\
V_N \\
\end{matrix}
\right) \, .
\end{equation}
\end{widetext}
The relative voltages $V_i/V$ can then be solved straightforwardly numerically using matrix inversion. The impedance $Z$ is solved using the relation
\begin{eqnarray}
I=\sum_i\frac{V-V_i}{Z_{{\rm C}i}} \, ,
\end{eqnarray}
which then leads to equation for impedance $Z$,
\begin{eqnarray}
\frac{1}{Z}\equiv\frac{I}{V}=\sum_i\left[ \frac{1}{Z_{{\rm C}i}}- \frac{1}{V}\frac{V_i}{Z_{{\rm C}i}} \right] \, .
\end{eqnarray}
As discussed in the main text, the total effective impedance includes also capacitors $C$ and $C_{\rm int}$. The answer for the total effective impedance is then
\begin{equation}
Z_{\rm eff}=\left( \mathrm{i}\omega C+ \mathrm{i}\omega C_{\rm int}+ Z^{-1} \right)^{-1} \, .
\end{equation}

\end{document}